\documentclass[10pt,journal]{IEEEtran}

\makeatletter
\def\ps@headings{\def\@oddhead{\mbox{}\scriptsize\rightmark \hfil \thepage}\def\@evenhead{\scriptsize\thepage \hfil \leftmark\mbox{}}\def\@oddfoot{}

\def\@evenfoot{}}
\makeatother

\newcommand{\noapp}[1]{{#1}}
\newcommand{\app}[1]{}
\newcommand{\SupplOf}{}
\newcommand{\SupplIn}{}

\usepackage{wrapfig}
\usepackage{times}
\usepackage{boxedminipage, epsfig,graphics,graphicx, multirow, array}

\usepackage{url}
\usepackage[dvipsnames]{xcolor}

\usepackage[vlined,ruled,linesnumbered]{algorithm2e}
\SetAlFnt{\footnotesize}

\SetCommentSty{mycommfont}
\SetKwComment{tcp}{$\triangleright$\ }{}
\SetKwComment{tcc}{//\ }{}
\SetKwInput{Require}{Require}
\usepackage{listings}

\usepackage[center]{caption}

\usepackage{acronym}

\usepackage{tabularx}
\usepackage{booktabs}
\usepackage{multicol}
\usepackage{multirow}
\usepackage{graphicx}

\usepackage{subfig}
\usepackage{comment}

\usepackage{fixltx2e}

\usepackage{upgreek}
\usepackage{amsmath,amssymb,amsthm,amsfonts,amsbsy}
\usepackage{mathtools}
\usepackage{bbm}

\usepackage{thmtools, thm-restate}
\declaretheorem[name=Proposition]{proposition}

\declaretheorem[name=Definition, sibling=proposition]{definition}

\SetKwRepeat{Do}{do}{while}

\DeclareMathOperator*{\argmax}{\arg\max}

\newcommand{\ALG}{{GREEDY}}
\newcommand{\alg}{{GREEDY}}

\newcommand{\ChanTrain}{{CHAN TRAIN}}

\newcommand{\GreedyRnd}{{GREEDY RND}}
\newcommand{\GreedyDeter}{{GREEDY DTR}}
\newcommand{\GreedyTrainRnd}{{GREEDY RND-SWT}}
\newcommand{\GreedyTrainDeter}{{GREEDY DTR-SWT}}

\newcommand{\OPTBTwo}{{OPT\textsubscript{B2}}}
\newcommand{\MDTOPT}{{MDTOPT}}

\newcommand{\evalFigWidth}{0.32}

\def\BibTeX{{\rm B\kern-.05em{\sc i\kern-.025em b}\kern-.08em
    T\kern-.1667em\lower.7ex\hbox{E}\kern-.125emX}}

\usepackage[latin1]{inputenc}

\usepackage[multiuser,nomargin,inline]{fixme}
\fxusetheme{color}
\fxsetup{targetlayout=colorcb}
\fxsetup{status=draft}
\FXRegisterAuthor{kmi}{ankmi}{\color{LimeGreen}kmi}
\FXRegisterAuthor{nko}{annko}{\color{Cyan}nko}
\FXRegisterAuthor{rem}{anrem}{\color{red}}

\graphicspath{{figures/}{figures/numericEval/}{figures/simuEval/}}

\begin{document}
\bstctlcite{IEEEexample:BSTcontrol}

\title{Greedy Multi-Channel Neighbor Discovery}

\author{Niels Karowski, Konstantin Miller and Adam Wolisz,~\IEEEmembership{Member,~IEEE}\IEEEcompsocitemizethanks{
\IEEEcompsocthanksitem N. Karowski and A. Wolisz are with the Telecommunication Networks Group (TKN) at the Technische Universit\"at Berlin, Germany.
Email: \{karowski,wolisz\}@tkn.tu-berlin.de, konstantin.miller@ieee.org
}\thanks{}
}

\maketitle

\begin{abstract}

The accelerating penetration of physical environments by objects with information processing and wireless communication capabilities requires approaches to find potential communication partners and discover services. 
In the present work, we focus on passive discovery approaches in multi-channel wireless networks based on overhearing periodic beacon transmissions of neighboring devices which are otherwise agnostic to the discovery process. 
We propose a family of low-complexity algorithms that generate listening schedules guaranteed to discover all neighbors. The presented approaches 
simultaneously depending on the beacon periods optimize the worst case discovery time, the mean discovery time, and the mean number of neighbors discovered until any arbitrary in time. 
The presented algorithms are fully compatible with technologies such as IEEE~802.11 and IEEE~802.15.4. 
Complementing the proposed low-complexity algorithms, we formulate the problem of computing discovery schedules that minimize the mean discovery time for arbitrary beacon periods as an integer linear problem. We study the performance of the proposed approaches analytically, by means of numerical experiments, and by extensively simulating them under realistic conditions. We observe that the generated listening schedules significantly -- by up to factor 4 for the mean discovery time, and by up to 300\% for the mean number of neighbors discovered until each point in time -- outperform the Passive Scan, a discovery approach defined in the IEEE 802.15.4 standard. Based on the gained insights, we discuss how the selection of the beacon periods influences the efficiency of the discovery process, and provide recommendations for the design of systems and protocols.
 \end{abstract}

\begin{IEEEkeywords}
Neighbor discovery, multi-channel, greedy algorithms.
\end{IEEEkeywords}

\section{Introduction}

We are currently observing a rapid augmentation of physical objects surrounding us with information processing and wireless communication capabilities. It is estimated that by 2020 25~\cite{gartnerIoT} up to 50~\cite{cisco_IoT} billion objects will be connected to the Internet. This development is leading us to a new era of computing. The resulting network of ``smart'' objects that interact with each other and exchange information without a direct human intervention, the so-called \ac{IoT}, will serve as a foundation for novel applications in a wide range of domains. 

In order to discover services of interest devices will need to detect other entities within communication range that are able to use common communication technology---the so called neighbors.

Neighbor discovery can be done in two fundamentally different ways. For an \emph{active} discovery, the discoverer broadcasts probe requests that must be answered by the neighbors.  
An active discovery is fast but has the drawback that all neighbors have to consume energy by (continuously) listening to potential inquiries even though they might only be interested in being detected but not in discovering their own neighborhood.
In contrast, \emph{passive} schemes perform the discovery by overhearing beaconing messages that are periodically broadcasted (with a specific \ac{BP}) by neighbors interested in being discovered.
The beaconing neighbors themselves are hereby agnostic to the discovery process. 
Let us emphasize that periodic beaconing is already used in many widely deployed technologies such as IEEE~802.11~\cite{ieee80211} and IEEE~802.15.4~\cite{ieee802154}.
In order to be compatible with current state-of-the-art technologies, such as IEEE~802.11 and IEEE~802.15.4, neighbor discovery must support multi-channel environments. 
Finally, we assume lack of time synchronization among the devices involved in the discovery process.

A frequently adopted objective for the design of discovery approaches is the minimization of the \ac{WDT}---time required to detect all \emph{potential} neighbors. A complete discovery is desirable, e.g., in order to avoid interference with neighbors when establishing a new network. In addition, minimizing the \ac{WDT} has the advantage of implicitly minimizing the consumed energy. 
Other applications are interested in the maximization of the number of discoveries until a given point in time, which we call the \ac{NDoT}, as, e.g., in the case of identifying potential forwarders in \aclp{DTN}.
Yet other applications benefit from discovering the individual neighbors as early as possible, e.g., emergency services. Their objective is thus the minimization of the \ac{MDT}. 
Since many devices in IoT environments will be battery powered, and will have limited computational resources, neighbor discovery should be performed in an energy-efficient way with low to moderate computational requirements.

This paper presents several novel contributions providing simple and efficient discovery algorithms applicable under realistic conditions:

We provide for the first time a full characterization of the class of listening schedules that are guaranteed to discover all neighbors (we call such schedules \emph{\textbf{complete}}), and that pointwise maximize the \ac{CDF} of the discovery times. The latter feature implies that they optimize all three mentioned performance metrics \emph{simultaneously}: \ac{WDT}, \ac{MDT}, and the \ac{NDoT}. We call these schedules \emph{\textbf{recursive}}, due to their specific structure.

Our second, practically most relevant, contribution consists of several approaches to construct listening schedules that, under certain assumptions, are recursive (and thus inherit the corresponding optimality properties). We define a family of low-complexity algorithms that we call \alg{}, due to their operation mode~\cite{West2001}. Further, we define an algorithm called CHAN TRAIN which is an extension of the \alg{} family, aiming at a reduction of the number of channel switches. 

In general, the performance of discovery algorithms is strongly dependent on the allowed set of \acp{BP}, 
i.e. periods with which beacons are transmitted. To this point - incompatible with the state-of-the-art wireless protocols -
assumptions about the beacon transmission patterns have been made so far in the literature. In this paper, we consider on one hand the most general case $\mathbb{F}_1$, a family containing all
possible \ac{BP} sets, but introduce in addition also two other practically important families of \ac{BP} sets---$\mathbb{F}_2$ and $\mathbb{F}_3$. They include all \ac{BP} sets supported by IEEE 802.15.4 and a large part
of the BI sets supported by IEEE 802.11---two widely adopted standards for wireless communication. 

We prove that for \ac{BP} sets from $\mathbb{F}_3$ the listening schedules computed by GREEDY and CHAN TRAIN are recursive (and thus complete, and optimal w.r.t. the three targeted performance metrics). Moreover, for \ac{BP} sets from $\mathbb{F}_2$ the computed schedules are complete and \ac{WDT}-optimal, while they are close-to-optimal w.r.t. the \ac{MDT}. Finally, we show that even for the most general case of $\mathbb{F}_1$ the computed schedule are complete, close-to-optimal w.r.t. the \ac{MDT}, and still within 30\%  of the optimum for the \ac{WDT}, while this gap decreases for an increasing number of channels.

Our third contribution demonstrates that even for arbitrary \ac{BP} sets from $\mathbb{F}_1$
complete and \ac{MDT}-optimal schedules are achievable, albeit only by solving an \ac{ILP}.
We prove that computed schedules are also \ac{WDT}-optimal for \ac{BP} sets from $\mathbb{F}_2$ and \ac{NDoT}-optimal for \ac{BP} sets from $\mathbb{F}_3$. This approach is attractive due to the broad range of supported \ac{BP} sets. However, it has a high computational complexity and memory consumption, restricting its usage to offline computations, and to scenarios with a moderate number of channels and size of the used \acp{BP}.

As additional contribution, we define an algorithm called \OPTBTwo{} that computes recursive schedules for scenarios, in which the cardinality of the \ac{BP} set is restricted to two entries. A summary of results is provided in Table~\ref{tab:strategy_overview}.

Unfortunately performing of such discovery in real multi-channel environments suffers under an implementation impact: non-negligible deaf periods occur during the execution of a channel switch resulting in potentially missing some beacons transmitted during such deaf periods. 
Due to this effect even algorithms provably generating complete schedules will, in reality, miss some neighbors - the percentage of missed neighbors can reasonably be expected to increase with the increase of the number of channel switches required by a given algorithm. 
In order to quantify this impact, we perform simulations using a realistic wireless model and device behavior expressing the results in the form of an additional performance metric --- the success rate, which is the fraction of neighbors discovered under this realistic conditions by any algorithm under consideration.
Using this additional performance metric we have derived our next contribution: We suggest two instances of the \ALG{} family of algorithms designed to reduce the number of channel switches and perform their simulative performance evaluation w.r.t achievable fraction of discovered neighbors.

In all evaluations, in addition to comparing the performance with the optimum, we perform a comparison against \acf{PSV}, a discovery scheme defined by the IEEE 802.15.4 standard. We observe that \ALG{} algorithms significantly (by up to several hundreds percent) outperform \ac{PSV} w.r.t. the \ac{MDT} and the \ac{NDoT} in all studied scenarios.

As our final contribution we discuss the strong impact the structure of allowed \ac{BP} sets has on the performance of discovery approaches, and provide recommendations for a \ac{BP} selection that supports efficient neighbor discovery. These recommendations may be useful, on the one hand, for the development of novel wireless communication based technologies that use periodic beaconing messages for management or synchronization purposes, and, on the other hand, for the \ac{BP} selection for existing technologies that support a wide range of \acp{BP}, such as IEEE~802.11.

The rest of this paper is structured as follows. 
Section~\ref{sec:related_work} discusses the related work.
Section~\ref{sec:system} describes the system definition and introduces the notation. 
Section~\ref{sec:perf_metrics} outlines the targeted performance metrics.
In Section~\ref{sec:BIFamilies} we present the identified families of \ac{BP} sets.
Section~\ref{sec:strategies} introduces recursive schedules, and our developed discovery algorithms, as well as \ac{PSV} that we use for comparison.
In Section~\ref{sec:perf_anal} we outline the analytical optimality results established for the proposed approaches.
In Sections~\ref{sec:num_exp} and~\ref{sec:simus}, we present the evaluation settings and results for numerical experiments and simulations, respectively.
In Section~\ref{sec:biSelection}, we provide recommendations on the selection of \acp{BP} supporting the discovery process. 
Finally, Section~\ref{sec:conclusion} concludes this paper and outlines future work.
 \section{Related Work}
\label{sec:related_work}

Most neighbor discovery approaches have in common that they divide the time into slots~\cite{Sun_2014_Survey}, and require each device to cooperate in the discovery process by being active - either transmitting or listening - in time slots following a pattern selected from a certain, more or less restrictive, set. 
Two devices discover each other if both are in complementary states in overlapping time slots, on the same channel. In contrast to many studies that focus on such mutual discovery, we focus on unilateral discovery, in which a device wants to discover some or all its neighbors by overhearing their regularly transmitted beacon messages not necessarily related specifically to any discovery process.
While many studies only have the objectives to guarantee a complete discovery, and to minimize the spent energy, we focus on optimizing the \ac{WDT}, the \ac{MDT}, and the \ac{NDoT}.

Discovery strategies can be classified into probabilistic and deterministic approaches, w.r.t. the selection of active states. 
Deterministic approaches work either with quorums, (co)prime numbers, or static/dynamic slot schemes. In quorum approaches discovery is based on the intersection of schedules generated either by a grid or a cyclic pattern~\cite{Lai_2010_Quorum,Tseng_2003_quorum}. In the former case time slots are arranged in a square matrix from which each device picks one column and one row to serve as its active slots. In the latter case discovery is achieved by constructing schedules based on cyclic difference sets which have guaranteed overlaps. With approaches based on (co)prime numbers, such as DISCO~\cite{dutta_2008_DISCO} and U-Connect~\cite{Kandhalu_2010_UConnect}, devices are active during time slots that are multiples of each device's selected prime number. Discovery is then guaranteed by the Chinese Remainder Theorem. Most recent work has been published in the area of the static/dynamic slot schemes. Discovery schedules generated by, e.g., Searchlight~\cite{Bakht_2012_Searchlight}, Hello~\cite{Wei_2014_Hello} and BlindDate~\cite{Wang_2015_Blinddate} consist of multiple cycles, where in each cycle there is one active slot at a fixed position and at least one dynamic slot whose position is shifted each cycle.
FlashLinQ~\cite{Wu_2013_FlashLinQ} is a PHY/MAC network architecture based on \ac{OFDM},  
which requires a strict synchronization between devices, which is difficult to achieve.
In~\cite{Cohen_2011_ContinuousNeighbor} devices perform a continuous collaborative discovery by forming a cluster after finding each other in order to reduce the expenses on each individual device to detect new neighbors.
Probabilistic approaches have in common that devices select their operational state out of at least two states, \emph{transmit} and \emph{listen}, with a predefined probability~\cite{mcglynn01,vasudevan09,Khalili10}. In~\cite{mcglynn01}, devices may also select with some probability an additional state \emph{sleep}. 

All the mentioned studies consider only single-channel environments, while most state-of-the-art wireless communication technologies allow devices to operate over multiple channels. In contrast, our proposed approaches consider multi-channel environments.
In addition, those studies require that the transmission pattern periodicities are coprime, or that instead of a single beacon message, a certain sequence is transmitted. The downside of these restrictions is that the network operators or service providers are no longer able to flexibly select beacon transmission patterns that are most appropriate for their targeted applications, used hardware, or the current operational state. For example, the required transmission patterns may interfere with sleeping patterns, which is particularly relevant for energy-constrained communication. Or, they may lead to a conflict with the deployed \ac{MAC} protocol, whose modification may be problematic due to, e.g., proprietary software, or protocols being implemented in hardware. In contrast, our approaches fully support state-of-the-art technologies such as, e.g., IEEE~802.11 and IEEE~802.15.4.

Discovery approaches supporting multiple channels have been mostly developed in the \ac{CR} context, in which a discovery is often termed rendezvous. Licensed spectrum owned by \acp{PU} is split up into multiple channels and an unused subset of these channels is then utilized by \acp{SU} for communication.
Typically, such approaches construct \ac{CH} sequences enabling \acp{SU} to discover each other without the use of common control channels. 
In~\cite{Theis_2011_RendezvousCR}, four \ac{CR} rendezvous approaches are described. The first is a probabilistic approach in which devices randomly select the operation state and channel.
The second is based on generated orthogonal sequences. Devices performing a rendezvous have to follow the same sequence and will eventually be active on the same channel and in the same time slot. The last two approaches use prime number modular arithmetic to guarantee rendezvous.
In~\cite{Zhang_2014_ETCH} two rendezvous protocols called ETCH are presented; SYNC-ETCH that requires global synchronization and creates \ac{CH} sequences using colored graphs and the asynchronous protocol ASYNC-ETCH.
ICH~\cite{Wu_2013_ICH} and ACH~\cite{Bian_2013_ACH} apply the static/dynamic slot scheme to multiple channels using cyclic quorum systems. 

Analogously to studies targeting single-channel settings, these studies require specific activity patterns that significantly restrict the potential of the devices to operate based on their state, operational goals, and data link layer technology.

In our previous work~\cite{Karowski11, Karowski13,willig10} we developed discovery approaches for IEEE~802.15.4 networks, in which beacons are sent periodically at \acp{BP} of the form $\tau\cdot 2^{BO}$, where $\tau$ is the duration of a superframe, and \acf{BO} is a parameter taking values between 0 and 14. In contrast, in the present work, we focus on efficient algorithms for broader families of \ac{BP} sets, even including arbitrary \ac{BP} sets that do not have any restrictions. By supporting a wide range of \ac{BP} sets we give the device or network operator the possibility to adapt the \ac{BP} to its specific requirements and the protocol stack, and ensure compatibility to existing technologies.

 
\section{System Definition and Notation}
\label{sec:system}

\newcommand{\cellheight}{1ex}
\begin{table*}[t]
\begin{center}
\begin{tabular}{m{6.4cm} m{11cm} @{}m{0pt}@{}}
\toprule
$\tau$ & Time slot duration &\\[\cellheight]
$C\subset\mathbb{N}^+$ & Set of channels &\\[\cellheight]
$\mathcal{L}\subset C\times\mathbb{N}$ & Listening schedule consisting of a sequence of (channel, time slot) pairs &\\[\cellheight]
$B\subset\mathbb{N}^+$ & Set of \acfp{BP} &\\[\cellheight]
$LCM(B)$ & Least common multiple of a set $B$ &\\[\cellheight]
$N$,~~$\nu \in N$ & Set of neighbors, one specific neighbor &\\[\cellheight]
$c_\nu \in C$,~~$b_\nu \in B$,~~$\delta_\nu \in \{0, \ldots, b_\nu - 1 \}$ & Operating channel, \ac{BP}, and beacon offset of neighbor $\nu$ &\\[\cellheight]
$\kappa=\left(c_\kappa,b_\kappa,\delta_\kappa\right)$,~~$\left(c_\nu,b_\nu,\delta_\nu\right)$ & Neighbor configuration $\kappa$ using \ac{BP} $b_\kappa$, channel $c_\kappa$ and offset $\delta_\kappa$; configuration of neighbor $\nu$ &\\[\cellheight]
$K_{BC}= \left\{\,\left(c,b,\delta\right)\mid c\in C, b\in B, \delta\in\{0,\ldots,b-1\}\,\right\}$ & Set of possible neighbor configurations for a \ac{BP} set $B$ and a set of channels $C$ &\\[\cellheight]
$K_c(t)=\left\{\,\left(c,b,t\bmod b\right)\mid b\in B\,\right\}$ & Set of configurations transmitting a beacon on channel $c$ in time slot $t$ &\\[\cellheight]
$\mathcal{T}_\nu=\left\{\delta_\nu+i \cdot b_\nu\right\}_{i\geq 0}$,~~$\mathcal{T}_\kappa=\left\{\delta_\kappa+i \cdot b_\kappa\right\}_{i\geq 0}$ & Beaconing time slots of neighbor $\nu$; beaconing time slots of a neighbor with configuration $\kappa$ &\\[\cellheight]
$\mathcal{B}_\nu=\left\{c_\nu\right\}\times\mathcal{T}_\nu$,~~$\mathcal{B}_\kappa=\left\{c_\kappa\right\}\times\mathcal{T}_\kappa$ & Beacon schedule of neighbor $\nu$; beacon schedule of a neighbor operating with configuration $\kappa$ &\\[\cellheight]
$T_\nu\left(\mathcal{L}\right)= \min\left\{\,t\in\mathcal{T}_\nu\mid\left(c_\nu,\,t\right)\in\mathcal{L}\,\right\}$ & Discovery time of neighbor $\nu$, given listening schedule $\mathcal{L}$ &\\[\cellheight]
$T_\kappa\left(\mathcal{L}\right)= \min\left\{\,t\in\mathcal{T}_\kappa\mid\left(c_\kappa,\,t\right)\in\mathcal{L}\,\right\}$ & Discovery time of all neighbors operating with configuration $\kappa$ &\\[\cellheight]
$P_{\kappa}, P_{c,b,\delta}$ & Probability that configuration $\kappa = (c,b,\delta)$ is selected by a neighbor &\\[\cellheight]
\bottomrule

\end{tabular}
\end{center}
\caption{Notation overview.}
\label{tab:system_model_parameter}
\end{table*}

We assume that each device has a single radio transceiver supporting the same transmission technology.
We consider a device (called in the following \emph{discoverer}) interested in detecting other devices (called \emph{neighbors} in the following) within its communication range.

We assume that each neighbor operates all the time on one of the channels from a set of channels $C$. 
We denote the set of neighbors by $N$ and the channel used by a neighbor $\nu\in N$ by $c_\nu\in C$.
Furthermore, we assume that each neighbor $\nu\in N$ announces its presence by periodically broadcasting beacon signaling messages every $b_\nu\tau$ seconds, where $b_\nu\in B$ is an integer called the \acf{BP}, and $\tau$ is a technology-dependent time unit. 
We also assume that the maximum beacon transmission time (time required to send one beacon) is smaller than $\tau$.

From the perspective of the discoverer time is divided into slots of length $\tau$ such that the $i$-th time slot contains the time period $\left[i\tau,\,(i+1)\tau\right)$, $i\in\mathbb{N}$. It is important to note that this definition of slotted time only reflects the view of the discoverer, therefore no synchronization between the discoverer and its neighbors is required.

Using this definition of time we denote the set of time slots when a neighbor $\nu$ sends a beacon by $\mathcal{T}_\nu$. Due to the periodicity of the beacon transmissions, $\mathcal{T}_\nu=\left\{i b_\nu+\delta_\nu\right\}_{i\in\mathbb{N}}$, where $\delta_\nu\in\left\{0,\ldots,b_\nu-1\right\}$. 
We call $\delta_\nu$ the offset of the neighbor $\nu$. Note that $\delta_\nu=t\bmod b_\nu$ for all $t\in\mathcal{T}_\nu$. We call the set of (channel, time slot) pairs given by $\mathcal{B}_\nu=\left\{c_\nu\right\}\times\mathcal{T}_\nu$ the beacon schedule of neighbor $\nu$.

With the introduced notation, each neighbor $\nu$ is represented by a tuple $\left(c_\nu,b_\nu,\delta_\nu\right)$, which we call a \textbf{neighbor configuration}. It is possible that multiple neighbors use the same configuration. Note that this does not necessarily lead to beacon collisions since the beacon transmission time is typically substantially smaller than $\tau$, and since the starting time of each neighbor is randomly distributed in the time slot $\delta_\nu$. For example, with IEEE~802.15.4, the default beacon size when operating in the 2.4 GHz frequency band is 38 symbols, as compared to the time slot duration $\tau \geq 960$ symbols.

For a given set of beacon periods $B$ and a set of channels $C$ we denote the set of possible neighbor configurations by $K_{BC}=\left\{\,(c,b,\delta)\mid c\in C,\,b\in B,\,\delta\in\{0,\ldots,b-1\}\,\right\}$. In analogy to $\mathcal{T}_\nu$ and $\mathcal{B}_\nu$, we define the beaconing time slots and the beacon schedule of a configuration $\kappa=\left(c_\kappa,b_\kappa,\delta_\kappa\right)\in K_{BC}$, and denote them with $\mathcal{T}_\kappa$ and $\mathcal{B}_\kappa$ respectively. 
In the following, the formulation that a configuration is sending its beacons during the time slot $t$, means that neighbors using this configuration are sending their beacons during time slot $t$.
We denote by $K_c(t)=\left\{\,\left(c,b,t\bmod b\right)\mid b\in B\,\right\}$ the set of configurations that send their beacons on channel $c$ during the time slot $t$.

We emphasize that we do not assume any coordination between the discoverer and the neighbors, or among the neighbors. In particular, we do not assume that the discoverer knows the individual \acp{BP} used in its neighborhood. However, we assume that the discoverer adopts a certain target \ac{BP} set $B$ to compute a listening schedule. $B$ may be the set of \emph{all} \acp{BP} permitted by the used communication technology, e.g. by the IEEE~802.15.4 standard. In some scenarios, however, the discoverer may be able to reduce the size of $B$ by the field of application, or to the values determined by a set of common policies, or to the values learned from past observations. In some cases, the discoverer may even only be interested in discovering neighbors using certain \acp{BP}, corresponding to certain applications or services, and deliberately use a set of \acp{BP} that does not contain all \acp{BP} potentially used by the neighbors.

In order to perform the discovery, the discoverer utilizes an algorithm that selects channels which are scanned during the individual time slots in order to search for beacons possibly transmitted by neighbors.
We call the resulting set of (channel, time slot) pairs a listening schedule, denoted by $\mathcal{L}\subset C\times\mathbb{N}$. Since we assume that the discoverer possesses a single radio transceiver and thus cannot simultaneously listen on multiple channels, we demand $c\neq c'\Rightarrow t\neq t'$ for all $\left(c,\,t\right),\left(c',\,t'\right)\in \mathcal{L}$. 
We denote the part of a listening schedule executed prior to a time slot $t$ by $\mathcal{L}_{t-1}=\left\{\,(c,t')\in\mathcal{L}\mid t'<t\,\right\}$, while $\mathcal{L}_{-1}=\emptyset$. 

To be able to generate optimized listening schedules, the discoverer requires assumptions about the probabilities $P_\kappa$ that a neighbor selects a certain configuration $\kappa = (c,b,\delta)$. In the following, we assume that the probability that a neighbor is using a certain channel is $1/{|C|}$ and the probability that a neighbor with a \ac{BP} $b$ is using an offset $\delta\in\{0,\ldots,b-1\}$ is $1/b$. We remark that we require no assumptions on the distribution of the \acp{BP}. 

\begin{figure}[t]
\centering
\includegraphics{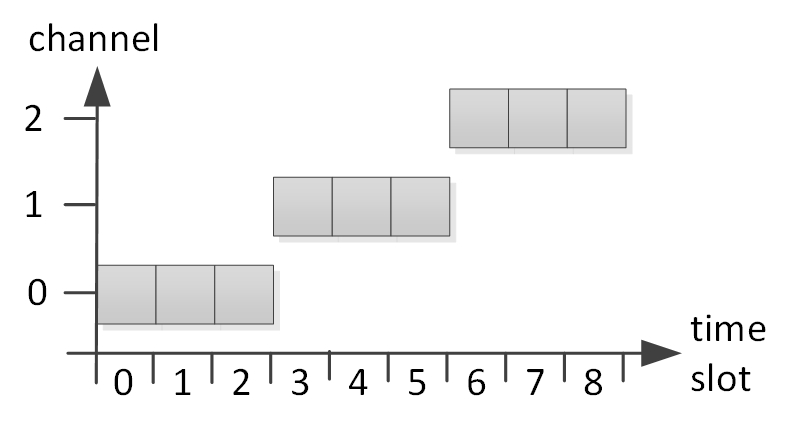}
\caption{Example of a listening schedule. Squares indicate the (channel, time slot) pairs scanned by the discoverer.}
\label{fig:ExampleSchedule}
\end{figure}

In this study, we focus on schedules that are guaranteed to discover all neighbors operating with any \ac{BP} $b \in B$ on any channel $c\in C$. We call such schedules \emph{complete}. More precisely, a complete schedule $\mathcal{L}$ is a schedule that contains at least one element from the beacon schedule $\mathcal{B}_\kappa$, for each configuration $\kappa\in K_{BC}$: $\mathcal{L}\cap\mathcal{B}_\kappa\neq\emptyset$, $\forall \kappa\in K_{BC}$. A complete schedule always exists. A simple example is a schedule that sequentially scans each channel for $\max\left(B\right)$ consecutive time slots, as depicted in Figure~\ref{fig:ExampleSchedule} for 3 channels and $\max(B)=3$. This strategy is defined in the IEEE~802.15.4 standard; we denote it as \acf{PSV} (see also Section~\ref{subsec:psv}). In the following, we only consider complete schedules.

For a schedule $\mathcal{L}$, we denote by $T_\nu(\mathcal{L})=\min\left\{\,t\in\mathcal{T}_\nu\mid(c_\nu,t)\in\mathcal{L}\,\right\}$ the discovery time of neighbor $\nu$. Similarly, we denote by $T_\kappa(\mathcal{L})=\min\left\{\,t\in\mathcal{T}_\kappa\mid(c_\kappa,t)\in\mathcal{L}\,\right\}$ the discovery time of a configuration $\kappa$. Whenever the considered schedule is clear from the context, we will simply write $T_\nu$ or $T_\kappa$, omitting the argument.

When designing the proposed discovery approaches, and analytically showing their optimality, we make the idealizing assumptions that there are no beacon losses due to collisions or interference, 
that the switching time between channels is equal to zero and a beacon transmission/reception time of zero.
However, later we will relax these assumptions when we evaluate the developed approaches by means of simulations under realistic conditions. In addition, in the present work, we assume that no neighbor enters or leaves the communication range of the discoverer during the discovery process. 

\section{Performance Metrics}
\label{sec:perf_metrics}

The following performance metrics will be used in order to assess the performance of neighbor discovery approaches: the \acf{NDoT}, the \acf{MDT}, the \acf{WDT}, the number of listening time slots (which is equivalent to the energy consumption), the number of channel switches, and the success rate. They are described in the following.

Applications which have delay constraints but needs to find as much neighbors as possible benefit from discovering the individual neighbors as early as possible.
This is achieved by a listening schedule that pointwise maximizes the \ac{CDF} of the discovery times, that is, when for each $t\geq 0$ there is no other listening schedule which has a higher value of the \ac{CDF} at $t$. The \ac{CDF} of discovery times can be interpreted as the expected fraction of discovered neighbors as a function of time. We call this metric the \textbf{\acf{NDoT}}, and we call a schedule that has a pointwise optimal \ac{CDF} of discovery times \ac{NDoT}-optimal. \ac{NDoT}-optimality implies that for each time $t$, the expected number of neighbors discovered prior to $t$ is optimal. We remark that not every setting admits a \ac{NDoT}-optimal schedule.

In order to make the \acp{CDF} of the discovery times comparable across different settings, we use discovery times $\tilde{T}_\kappa(\mathcal{L})=\frac{T_\kappa(\mathcal{L})}{\max(B)|C|}\in[0,\infty)$ normalized to the minimum time 
required to discover all potential neighbors.

A weaker performance metrics that considers the individual discovery times is the \textbf{\acf{MDT}}, given by $\sum_{\kappa\in K_{BC}} P_\kappa T_\kappa(\mathcal{L})$. We remark that \ac{NDoT}-optimality implies \ac{MDT}-optimality. A \ac{MDT}-optimal schedule exists in any setting. 

For a \ac{BP} set $B$, a set of channels $C$, and a listening schedule $\mathcal{L}$, the \textbf{\acf{WDT}} is the number of time slots until all \emph{potential} neighbors are discovered. It is given by $\max_{\kappa\in K_{BC}} T_\kappa(\mathcal{L})$. The optimum \ac{WDT} is given by $\max(B)|C|$, for arbitrary \ac{BP} sets and channel sets (see Section~\app{\ref{sec:opt_cdt} \SupplOf{}}\noapp{3 in~\cite{Karowski18_tech_report}} for a proof). We remark that a \ac{NDoT}-optimal schedule is also \ac{WDT}-optimal.

The amount of energy required to execute the schedule can be expressed by the \textbf{number of listening time slots}. The number of listening time slots is always less or equal than the \ac{WDT}. In particular, it may be strictly smaller than the \ac{WDT} due to the fact that a listening schedule may contain idle time slots during which no scan is performed. However, \ac{WDT}-optimal schedules are also optimal w.r.t.\@ the number of listening time slots (and thus w.r.t.\@ the energy consumption), as shown in Section~\app{\ref{sec:opt_cdt} \SupplIn{}}\noapp{3 in~\cite{Karowski18_tech_report}}.

The \textbf{number of channel switches} is the number of times a device has to change the listening channel when executing a schedule. The motivation for considering this metric stems from the fact that when a discoverer performs a channel switch it is in a deaf period in which it is not able to receive any messages, which may lead to losing beacons. Thus, a schedule with less switches is preferable.

Finally, in practical deployment, even a schedule which is complete under the previously stated idealizing assumptions may fail to discover all neighbors after its first execution. Possible reasons include beacon collisions and deaf periods due to channel switches. We call the fraction of discovered neighbors in a given environment the \textbf{success rate}.

\section{Considered Families of \ac{BP} Sets}
\label{sec:BIFamilies}

\begin{table*}[t]
\begin{center}
\begin{tabular}{>{\arraybackslash}m{1.7cm} m{14cm} @{}m{0pt}@{}}
\toprule
$\mathbb{F}_1$ & This is the most general family of \ac{BP} sets that contains any finite subset of $\mathbb{N}^+$. &\\[1.5ex] \midrule
$\mathbb{F}_2$ & Family $\mathbb{F}_2\subset\mathbb{F}_1$ includes all \ac{BP} sets $B$ that contain a multiple of their \acp{LCM}: \newline $\max(B)=LCM(B)$. &\\[1.5ex] \midrule
$\mathbb{F}_3$ & Family $\mathbb{F}_3\subset\mathbb{F}_2$ includes all \ac{BP} sets $B$ in which each element is an integer multiple of each of the smaller elements: $\max(B')=LCM(B')$, $\forall B'\subseteq B$. &\\[3ex] \midrule
$\mathbb{F}_4$ & Family $\mathbb{F}_4\subset\mathbb{F}_3$ includes all \ac{BP} sets $B=\left\{b_0,\ldots,b_{n-1}\right\}$ whose elements are powers of the same base, potentially multiplied with a common coefficient: $\exists\,k,c\in\mathbb{N}^+$ and $\exists\,e_0,\ldots,e_{n-1}\in\mathbb{N}$ such that $b_i=kc^{e_i}$, $\forall i\in\{0,\ldots,n-1\}$. &\\[3ex] \midrule
$\mathbb{F}_{\textrm{IEEE 802.15.4}}$ & This family contains all \ac{BP} sets $B$ defined by the IEEE~802.15.4 standard. All elements $b\in B$ must have the form $b=2^{BO}$, where $BO$ is a network parameter called the beacon order that can be assigned a value between 0 and 14. Possible periods lie in the range between approx. 15.36 $ms$ and 252 $s$. &\\[6ex] \midrule
$\mathbb{F}_{\textrm{IEEE 802.11}}$ & The family of \ac{BP} sets $B$ allowed by the IEEE~802.11 standard contains arbitrary sets such that each element $b\in B$ can be represented by a 16 bit field, that is, $b\in\left\{1,\ldots,2^{16}-1\right\}$. &\\[3.5ex]
\bottomrule
\end{tabular}
\end{center}
\caption{Characterization of identified families of \ac{BP} sets.}
\label{tab:sets}
\end{table*}

\begin{figure}[t]
\centering
\includegraphics[width=0.4\textwidth]{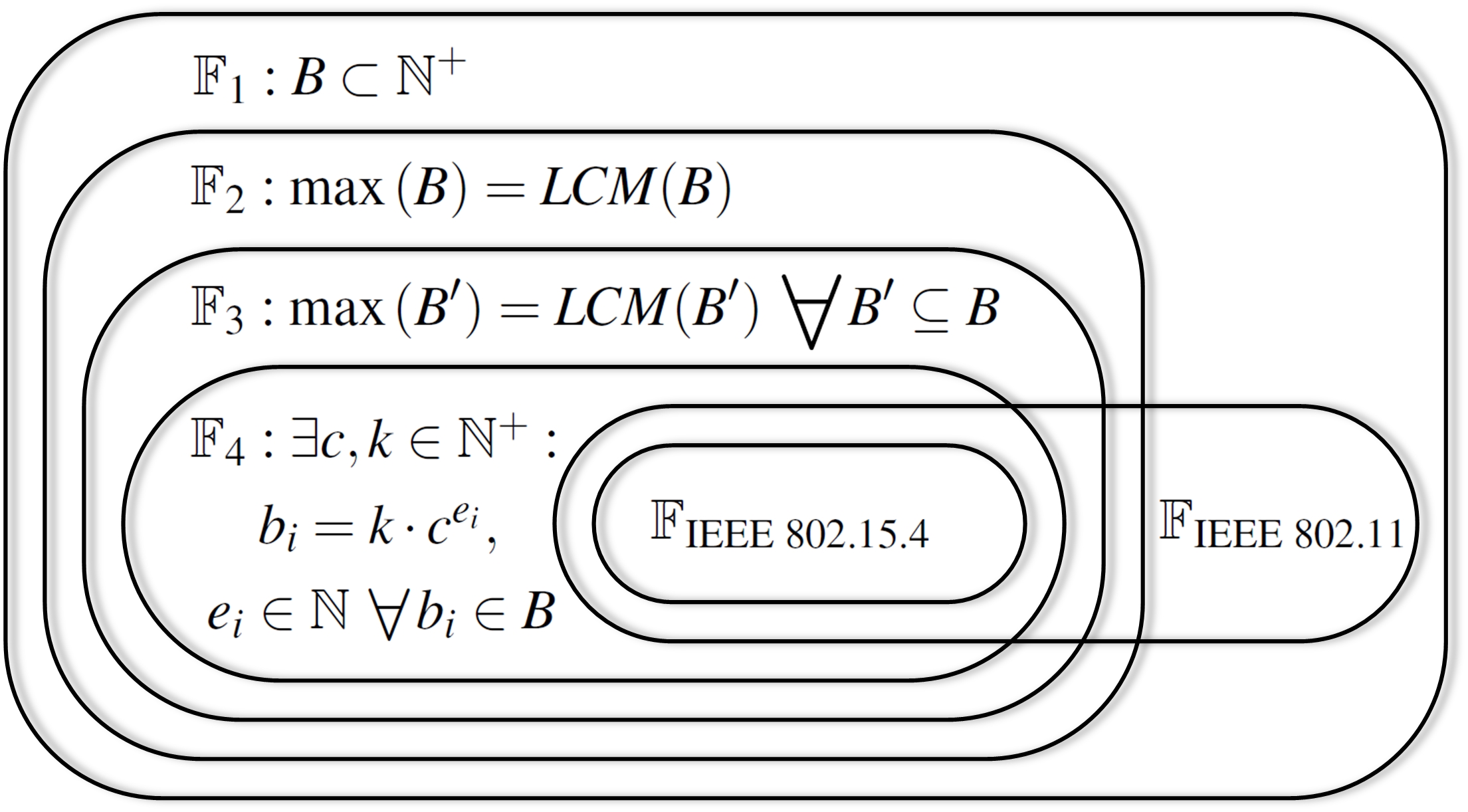}
\caption{Hierarchy of families of \ac{BP} sets.}
\label{fig:BIFamilies}
\end{figure}

In our study, we focus on three families of \ac{BP} sets: $\mathbb{F}_1$, $\mathbb{F}_2$, and $\mathbb{F}_3$, with $\mathbb{F}_1\supset\mathbb{F}_2\supset\mathbb{F}_3$. 

\begin{itemize}
	\item The family $\mathbb{F}_1$ contains any finite subset of $\mathbb{N}^+$.
	\item The family $\mathbb{F}_2$ contains sets $B$ in which all elements are proper divisors of the largest \ac{BP} $max(B)$, i.e. $max(B)=\text{\ac{LCM}}$, such as, for example, $\{x, y, z, xyz\}$, for arbitrary $x, y, z \in \mathbb{N}^+$. 
  \item The family $\mathbb{F}_3\subset\mathbb{F}_2$ contains sets $B$ in which each element is an integer multiple of each of the smaller elements, i.e. subsets of $B$ are in $\mathbb{F}_2$. Examples are $\{x, xy, xyz\}$, or $\{2^x, 2^y, 2^z\}$, for arbitrary $x, y, z \in  \mathbb{N}^+$. 
	\end{itemize}

In addition, we denote by $\mathbb{F}_4$ a generalization of the family of \ac{BP} sets $\mathbb{F}_{\textrm{IEEE 802.15.4}}$, which is defined by the IEEE~802.15.4 standard.
While $\mathbb{F}_1$ is the most general family which contains all \ac{BP} sets, $\mathbb{F}_2$ and $\mathbb{F}_3$ are of particular interest since they support fast discovery with low complexity. Notably, both $\mathbb{F}_2$ and $\mathbb{F}_3$ completely include the \ac{BP} sets supported by IEEE~802.15.4 and a large part of the \ac{BP} sets supported by IEEE~802.11. 
The overview of the hierarchy of families of \ac{BP} sets is depicted in Figure~\ref{fig:BIFamilies}. The definitions of the individual families are provided in Table~\ref{tab:sets}. 

In the following, w.l.o.g., we only consider \ac{BP} sets whose \ac{GCD} is 1 since a listening schedule for a \ac{BP} set with \ac{GCD} $d\neq 1$ is equivalent to a listening schedule based on the transformed set $B$ in which each element is divided by $d$, and the time slot duration $\tau$ is substituted by $\tau'=\tau d$. This transformation allows to reduce the computational complexity, which is particularly important for the \ac{ILP}-based approaches. For example, instead of the set $B =\{4, 10\}$, we may consider the set $\{2,5\}$, assuming $\tau'=2\tau$.

 \section{Optimized Discovery Strategies}
\label{sec:strategies}

In this section, we first characterize the class of recursive listening schedules. We then describe the proposed algorithms for computing optimized listening schedules. Finally, we introduce \ac{PSV}, a discovery approach defined by IEEE~802.15.4 that we use for a comparative performance evaluation.

\subsection{Recursive Listening Schedules}
\label{sec:recursive}

We define a recursive schedule as a schedule that discovers the neighbor configurations in the order of their \acp{BP}.

\begin{definition}[Recursive schedule]
\label{def:recursive_schedule}
For a \ac{BP} set $B$, and a set of channels $C$, a schedule is called recursive if and only if all configurations with a \ac{BP} $b\in B$ are discovered during the first $b|C|$ time slots.
\end{definition}

Note that this definition implies that a recursive schedule for a set $B$ contains a recursive schedule for a \ac{BP} set $B'\subset B$, justifying the naming.

Recursive schedules have a very compelling property -- they are equivalent to the class of schedules that are complete, and optimal w.r.t. the \ac{WDT}, \ac{MDT}, and \ac{NDoT}. 
The proof is presented in Section~\app{\ref{app:recursive} \SupplOf{}}\noapp{4 in~\cite{Karowski18_tech_report}}.

\subsection{\ALG{} Computation of Efficient Listening Schedules}
\label{sec:greedy}

An algorithm belongs to the class of \ALG{} algorithms if in each time slot it scans a channel that maximizes the expected number of discoveries. Since there may exist several such channels, \ALG{} is not a single algorithm but a family of algorithms that differ in the \textbf{tiebreaker rule} that selects one channel from a set of candidates. This definition is constructive and can be turned into a practical implementation in a straightforward manner. It is formalized in Definition~\ref{def:mindy}.

\begin{restatable}[\ALG{}]{definition}{mindy}
\label{def:mindy}
An algorithm $A$ is in \ALG{} if for a \ac{BP} set $B\in\mathbb{F}_1$ and a channel set $C$, in every time slot $t$ $A$ scans a channel $c_t\in C$ such that
\begin{equation}
c_t\in\argmax_{c\in C} \sum_{\kappa \in K_c(t)\,:\,\mathcal{B}_\kappa \cap \mathcal{L}_{t-1}=\emptyset} P_\kappa \,.
\end{equation}
\end{restatable}

The pseudo code describing the operation of \ALG{} algorithms is presented in Algorithm~\ref{alg:greedy}. As input, the algorithm obtains the set of channels $C$, the set of \acp{BP} $B$, and the configuration probabilities $P_\kappa$. It proceeds by iterating over time slots until all possible configurations (based on $C$ and $B$) have been considered. 
For each time slot, it first computes the expected fraction of neighbors that can be discovered by scanning each of the channels. It is equal to the sum of probabilities $P_\kappa$ for configurations that send their beacons in the given time slot on the given channel, that have not been considered previously. This computation is presented separately in Algorithm~\ref{alg:numConfs}. If this sum is 0 for all channels, the time slot remains idle. Otherwise, a set of candidates is formed by selecting those channels that maximize the expected number of discoveries. Then, the tiebreaker rule is used to select a particular channel from the set of candidates. We remark that a tiebreaker rule may require access to $C$, $B$, $P$, $\mathcal{L}$, or $t$. Finally, the schedule is updated and the algorithm proceed to the next time slot.

A \ALG{} algorithm terminates when all possible configurations have been covered, and thus all neighbors will be discovered when executing the generated listening schedule.
For \ac{BP} sets from $\mathbb{F}_2$, the \ac{WDT}-optimality of \ALG{} algorithms implies that only time slots $\{0,\ldots,\max(B)|C|-1\}$ need to be scanned (see Section~\ref{sec:perf_anal} for details). For \ac{BP} sets from $\mathbb{F}_1\setminus\mathbb{F}_2$, the worst-case upper bound on the runtime is $LCM(B)|C|$ time slots (the proof is similar to the proof of Proposition~\app{\ref{prop:greedy_f2} \SupplIn{}}\noapp{6 in~\cite{Karowski18_tech_report}} and is omitted for brevity). The performance evaluation, however, revealed that runtime typically lies between $\max(B)|C|$ and $2\max(B)|C|$ time slots. The presented pseudo code also allows an online execution.

Individual instances of the \ALG{} class are defined by the tiebreaker rule (line~\ref{alg:line_tiebreaker} of Algorithm~\ref{alg:greedy}) which describes the selection of a channel to be scanned next from the set of candidates $C_{\text{candidates}}$.
We will consider two deterministic and two probabilistic rules, described in the following.

\vspace{0.1cm}
\textbf{\GreedyRnd{}} randomly selects a channel from $C_{\text{candidates}}$.

\vspace{0.1cm}
\textbf{\GreedyDeter{}} selects the channel with the highest channel identifier from $C_{\text{candidates}}$.

\vspace{0.1cm}
\textbf{\GreedyTrainRnd{}} tests if the channel scanned in the previous time slot is in $C_{\text{candidates}}$. If yes, it is selected. If no, it proceeds as \GreedyRnd{}. By prioritizing the most recently selected channel, \GreedyTrainRnd{} tries to reduce the number of channel switches.

\vspace{0.1cm}
\textbf{\GreedyTrainDeter{}} is similar to \GreedyTrainRnd{} but without a random component. It tests if the channel scanned in the previous time slot is in $C_{\text{candidates}}$. If yes, it is selected. If no, it proceeds as \GreedyDeter{}.

\begin{algorithm}[t]
\DontPrintSemicolon
\SetArgSty{textup}
\SetInd{0.5em}{1.0em}
\caption{\textsc{GREEDY discovery algorithms}}
\label{alg:greedy}
\KwIn{$C$, $B$ 													\tcp*{\parbox[t]{.40\linewidth}{Set of channels, set of \acp{BP}}}}
\KwIn{$P = (P_{\kappa},\kappa \in K_{BC})$ \tcp*{\parbox[t]{.40\linewidth}{Configuration probabilities}}}
\KwOut{$\mathcal{L}$ \tcp*{\parbox[t]{.40\linewidth}{Listening schedule}}}
$t \gets 0$\;
$\mathcal{L} \gets \emptyset$\;
\While{$\exists\, \kappa \in K_{BC}\; \text{with}\; \mathcal{B}_{\kappa} \cap \mathcal{L} = \emptyset$}{
	${\text{discProbs} \gets \text{DiscProbs}(C,B,P,\mathcal{L},t)}$ \tcp*{\parbox[t]{.22\linewidth}{Returns array}}
	\If(\tcp*[f]{\parbox[t]{.22\linewidth}{Time slot not idle}}){$\max(\text{discProbs}) > 0$}{
			$C_{\text{candidates}}\leftarrow\argmax (\text{discProbs})$ \tcp*{\parbox[t]{.22\linewidth}{Returns set}}
			$c_t\leftarrow\text{tiebreak}\left(C_{\text{candidates}}\right)$\;\label{alg:line_tiebreaker}
			$\mathcal{L} \gets \mathcal{L} \cup \{(c_t,t)\}$\;
	}
	$t \gets t +1 $\;
}
\end{algorithm}

\subsection{CHAN TRAIN -- Reducing the Number of Channel Switches}
\label{sec:chantrain}

In this section, we propose an algorithm named~\ChanTrain{}, which aims at heuristically reducing the number of channel switches.  In contrast to \ALG{}, \ChanTrain{} will stay on a selected channel if subsequent time slots result in at least the same sum of discovery probabilities $P_{\kappa}$ as the previous time slot considering configurations that have not yet been covered which may result in a non-\ALG{} behavior. The pseudo code for CHAN TRAIN is presented in Algorithm~\ref{alg:chantrain}.

Analogously to GREEDY, CHAN TRAIN first computes the expected fraction of neighbors that can be discovered by scanning each of the channels. If none of the channels admits a discovery, the time slot remains idle. Otherwise, a set of candidates is formed by selecting channels that maximize the expected number of discoveries. Out of those, CHAN TRAIN selects the channel which maximizes the sum of two values: (i) the number of consecutive previous time slots allocated on this channel (which is non-zero for the channel scanned during the time slot $t-1$, and 0 for all other channels), and (ii) the number $t'$ of consecutive time slots starting with time slot $t$ with at least the same expected number of discoveries as during the time slot $t$. 
It then jumps to the time slot $t+t'$ and repeats the procedure. If multiple channels maximize the consecutive number of time slots that can be scanned in sequence, the one with the lowest identifier is selected. Note, however, that also other tiebreaker rules may be deployed here. 

Note that for \ac{BP} sets from $\mathbb{F}_3$, CHAN TRAIN belongs to the family GREEDY (see Section~\app{\ref{sec:optim_chantrain} \SupplOf{}}\noapp{6 in~\cite{Karowski18_tech_report}} for a proof). However, for \ac{BP} sets from $\mathbb{F}_1\setminus\mathbb{F}_3$ \ChanTrain{} is not necessarily GREEDY, due to the fact that it may jump over multiple time slots, in which a different channel may maximize the expected number of discoveries, as illustrated in Example~\app{\ref{exa:chantrain_nongreedy} \SupplIn{}}\noapp{3 in~\cite{Karowski18_tech_report}}. The overall complexity of CHAN TRAIN is higher than that of the \ALG{} algorithms due to the additional computation of the maximum number of consecutive time slots a device may stay on a channel.

\begin{algorithm}[t]
\DontPrintSemicolon
\SetArgSty{textup}
\SetInd{0.5em}{1.0em}
\caption{\textsc{CHAN TRAIN}}
\label{alg:chantrain}
\KwIn{$C$, $B$ \tcp*{\parbox[t]{.40\linewidth}{Set of channels, set of \acp{BP}}}}
\KwIn{$P = (P_{\kappa},\kappa \in K_{BC})$ \tcp*{\parbox[t]{.40\linewidth}{Configuration probabilities}}}
\KwOut{$\mathcal{L}$ \tcp*{\parbox[t]{.40\linewidth}{Listening schedule}}}
$t \gets 0$\;
$\mathcal{L} \gets \emptyset$\;
\While{$\exists\, \kappa \in K_{BC}\; \text{with}\; \mathcal{B}_{\kappa} \cap \mathcal{L} = \emptyset$}{
	${\text{discProbs} \gets \text{DiscProbs}(C,B,P,\mathcal{L},t)}$ \tcp*{\parbox[t]{.22\linewidth}{Returns array}}
	\If(\tcp*[f]{\parbox[t]{.22\linewidth}{Skip idle slot}}){$\max(\text{discProbs}) = 0$}{
		$t \gets t + 1$\;
		$\text{\textbf{continue}}$\;
	}
    $C_{\text{candidates}}\gets \argmax (\text{discProbs})$ \tcp*{\parbox[t]{.22\linewidth}{Returns set}}
	\ForEach{$c \in C_{\text{candidates}}$}{
		$t' \gets t$\;
		$\mathcal{L}' \gets \mathcal{L}$\;
		\Do(\tcp*[f]{\parbox[t]{.36\linewidth}{Determine future chan train}}){$\text{DiscProbs}(\{c\},B,P,\mathcal{L}',t') \geq \max(\text{discProbs})$}{
			$\mathcal{L}' \gets \mathcal{L}' \cap \{(c,t')\}$\;
			$t' \gets t' + 1$\;
		}
		$t_f[c] \gets t' - t$\;
		$t' \gets t -1$\;
		\While(\tcp*[f]{\parbox[t]{.34\linewidth}{Determine past chan train}}){$(c,t') \in \mathcal{L}$}{
			$t' \gets t' - 1$\;
		} 
		$t_{train}[c] \gets t_f[c] + (t - t' - 1)$ \tcp*{\parbox[t]{.22\linewidth}{Total train length}}
	}
	$c_{train} \gets \min(\argmax(t_{train}))$ \tcp*{\parbox[t]{.22\linewidth}{Selected channel}}
	$\mathcal{L} \gets \mathcal{L}\cup\{(c_{train},t')\}_{t' \in \{t,\ldots,t + t_f[c_{train}] - 1\}}$\;
	$t \gets t + t_f[c_{train}]$\;
}
\end{algorithm}

\subsection{\MDTOPT{} -- Minimizing \ac{MDT} for Arbitrary \ac{BP} Sets}
\label{sec:genopt}

The low-complexity \ALG{} algorithms presented so far may fail to achieve \ac{MDT}-optimality for \ac{BP} sets from $\mathbb{F}_1\setminus\mathbb{F}_3$, as illustrated in Example~\app{\ref{exa:greedy_suboptimal} \SupplIn{}}\noapp{2 in~\cite{Karowski18_tech_report}}, and described in more details in Section~\ref{sec:perf_anal} (though they are still very efficient or close-to-optimal even for these \ac{BP} sets, as observed from the evaluation results). In this section, we formulate an \acf{ILP} which we call \MDTOPT{} that minimizes the \ac{MDT} for arbitrary \ac{BP} sets from $\mathbb{F}_1$. To formulate \MDTOPT{}, we define the following optimization variables.

\begin{equation*}
x_{ctb} = \begin{cases}
				\begin{aligned}
		   \textrm{1  ,} & \textrm{ if configuration } \left(c,b,t\bmod b\right) \textrm{ is detected} \\
										& \textrm{ during scan of channel } c \textrm{ in time slot } t \\
           \textrm{0  ,} & \textrm{ otherwise}  
				\end{aligned}					
        \end{cases}
\end{equation*}

\begin{equation*}
h_{ct} = \begin{cases}
				\begin{aligned}
           \textrm{1  ,} & \textrm{ if channel } c \textrm{ is scanned during time slot } t \\
           \textrm{0  ,} & \textrm{ otherwise}  
				\end{aligned}	
        \end{cases}
\end{equation*}

\begin{algorithm}[t]
\DontPrintSemicolon
\SetInd{0.5em}{1.0em}
\caption{DiscProbs}
\label{alg:numConfs}
\KwIn{$C$, $B$ \tcp*{\parbox[t]{.57\linewidth}{Set of channels, set of \acp{BP}}}}
\KwIn{$(P_{\kappa},\kappa \in K_{BC})$ \tcp*{\parbox[t]{.57\linewidth}{Configuration probabilities}}}
\KwIn{$\mathcal{L}_{t-1}$ \tcp*{\parbox[t]{.57\linewidth}{Previously scanned (channel, time slot) pairs}}}
\KwIn{$t$ \tcp*{\parbox[t]{.57\linewidth}{Current time slot $t$}}}
\KwOut{$\text{discProbs}$ \tcp*{\parbox[t]{.57\linewidth}{Discovery probability per channel}}}
\ForEach{$c\in C$}{
	$\text{discProbs}[c]\leftarrow 0$\;
	\ForEach{$b\in B$}{
		\If{$\mathcal{B}_{(c,b,t\bmod b)} \cap \mathcal{L}_{t-1} = \emptyset$}{
			$\text{discProbs}[c] \gets \text{discProbs}[c] + P_{(c,b,t\bmod b)}$\;
		}
	}
}
\end{algorithm}

Now, \MDTOPT{} can be formulated as follows.

\begin{align*}
\text{min}\quad & \sum_{c\in C}\sum_{b\in B}\sum_{t=0}^{LCM(B)|C|-1}{x_{ctb} P_{(c,b,t \bmod b)} t}  \notag   \\
\text{s.t.}\quad &\sum_{i=0}^{\frac{LCM(B)  \left|C\right|}{b} - 1}{x_{c,ib+\delta,b}}=1 \label{eq:C1}\tag{C1} \\
&\qquad \text{for all}\;c\in C, b\in B,\;\delta\in\left\{0,\ldots,b-1\right\} \\
&x_{ctb}\leq h_{ct}\label{eq:C2} \tag{C2}\\
&\qquad \text{for all}\;c\in C,b\in B,t\in\left\{0,\ldots,LCM(B)|C|-1\right\}  \\
&\sum_{c\in C}h_{ct}\leq 1 \label{eq:C3} \tag{C3} \\
&\qquad \text{for all}\;t\in\left\{0,\ldots,LCM(B)|C|-1\right\}\,.
\end{align*}

In this formulation, constraint~\eqref{eq:C1} ensures that each configuration is detected,~\eqref{eq:C2} ensures that a configuration $\left(c,b,t\bmod b\right)$ can only be detected if channel $c$ is scanned during time slot $t$,~\eqref{eq:C3} makes sure that at most one channel is scanned during a time slot. Note that constraints \eqref{eq:C1} - \eqref{eq:C3} describe a generic listening schedule that can be used with alternative objective functions to compute schedules that are optimal w.r.t.\@ different targeted performance metrics.

We remark that it is necessary and sufficient for the optimization to consider $LCM(B)|C|$ time slots, since this is the worst case for \ac{MDT}-optimal schedules (see Section~\app{\ref{app:mdt_opt_cdt_opt} \SupplOf{}}\noapp{8 in~\cite{Karowski18_tech_report}} for a proof). We remark that \MDTOPT{} computes \ac{MDT}-optimal schedule for any probability distribution $\left(P_\kappa,\kappa\in K_{BC}\right)$. The assumptions of a uniform distribution over channels and beacon offsets are not required.

\subsection{\OPTBTwo{} - Special case: $\lvert B\rvert=2$}
\label{sec:opt2}

\begin{figure}[t]
\centering
\includegraphics[width=0.45\textwidth]{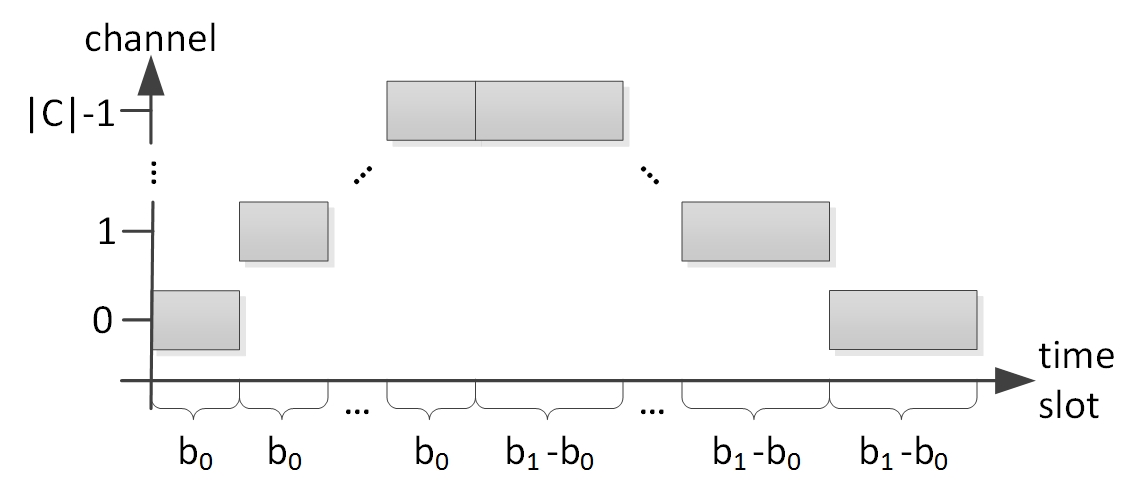}
\caption{Structure of listening schedules created by \OPTBTwo{}.}
\label{fig:OPTB2}
\end{figure}

For the special case of \ac{BP} sets containing exactly two elements, recursive listening schedules can be computed for arbitrary \ac{BP} values as follows (see Figure~\ref{fig:OPTB2} for an illustration).

\begin{restatable}[\OPTBTwo{}]{definition}{opt2}
\label{def:opt2}
For a \ac{BP} set $B=\left\{b_0,\,b_1\right\}$, with $b_0<b_1$, and a set of channels $C$, \OPTBTwo{} scans channel $j\in\{0,\ldots,{|C|-1}\}$ during the time slots $\left\{jb_0,\ldots,(j+1)b_0-1\right\}$ and 
\begin{gather*}
\left\{|C|b_0+(|C|-j-1)(b_1-b_0),\ldots,\right.\\
\left.|C|b_0+(|C|-j)(b_1-b_0)-1\right\}\,.
\end{gather*}
\end{restatable}

\subsection{Passive Scan in IEEE~802.15.4}
\label{subsec:psv}

The IEEE~802.15.4 standard defines four types of scanning techniques~\cite{ieee802154}. 
Since in our work we focus on passive discovery techniques, we compare our approaches against one of these scanning techniques, the passive scan denoted by PSV, in which the discoverer only listens to beacon messages. 
PSV proceeds by sequentially listening on each channel $c\in C$ for $\max(B)$ time slots. Thus, channel $c_j$, with $j\in\{0,\ldots,|C|-1\}$, is scanned during the time slots $\{j\max(B),\ldots,(j+1)\max(B) - 1\}$.

\section{Analytical Performance Evaluation}
\label{sec:perf_anal}

In this section we outline the analytical performance results for the proposed approaches. For a better readability and due to the space constraints, their rigorous formulations and proofs are presented~\app{\SupplIn{}}\noapp{in~\cite{Karowski18_tech_report}}. A summary is provided in the end of this section and in Table~\ref{tab:strategy_overview}.

\renewcommand{\cellheight}{1ex}
\begin{table*}[t]
\begin{center}
\begin{tabular}{>{\arraybackslash}m{1.9cm} >{\centering\arraybackslash}m{1.9cm} >{\centering\arraybackslash}m{1.9cm} >{\centering\arraybackslash}m{1.9cm} >{\centering\arraybackslash}m{1.9cm} >{\centering\arraybackslash}m{2.3cm} >{\centering\arraybackslash}m{3cm} @{}m{0pt}@{}}
\toprule
\textbf{Strategy}	 	& \textbf{Completeness} & \textbf{\ac{WDT} optimality} 	& \textbf{\ac{MDT} optimality} 	& \textbf{\ac{NDoT} optimality} & \textbf{Channel switches optimality} & \textbf{Complexity} &\\[3ex] \toprule
\ALG{} 							& $\mathbb{F}_1$ 				& $\mathbb{F}_2$ 							 	& $\mathbb{F}_3$ 						  	& $\mathbb{F}_3$ 								& 																		 & $\mathcal{O}\left(\left|C\right|^2\left|B\right|LCM(B)\right)$ &\\[\cellheight]
CHAN TRAIN 					& $\mathbb{F}_1$				& $\mathbb{F}_2$ 								& $\mathbb{F}_3$ 								& $\mathbb{F}_3$  							& 																		 & $\mathcal{O}\left(\left|C\right|^2\left|B\right|LCM(B)^2\right)$ &\\[\cellheight]
\MDTOPT{} 					& $\mathbb{F}_1$				& $\mathbb{F}_2$ 								& $\mathbb{F}_1$ 								& $\mathbb{F}_3$ 								& & NP-hard &\\[\cellheight]
\OPTBTwo{}					& $\mathbb{F}_1$	 			& $\mathbb{F}_1\;\left(|B|=2\right)$ & $\mathbb{F}_1\;\left(|B|=2\right)$ & $\mathbb{F}_1\;\left(|B|=2\right)$ & & $\mathcal{O}\left(C\right)$ &\\[\cellheight]
PSV 								& $\mathbb{F}_1$		  	& $\mathbb{F}_1$ 								& 															& 															& $\mathbb{F}_1$ & $\mathcal{O}\left(C\right)$ &\\[\cellheight]
(SW)OPT 						& $\mathbb{F}_3$			  & $\mathbb{F}_3$ 								& $\mathbb{F}_3$ 								& $\mathbb{F}_3$ 								& & NP-hard &\\[\cellheight]
\bottomrule
\end{tabular}
\end{center}
\caption{Optimality and complexity results overview. Note that for $\mathbb{F}_2$ and $\mathbb{F}_3$, $LCM(B)$ reduces to $\max(B)$.}
\label{tab:strategy_overview}
\end{table*}

\subsection{Performance of \ALG{} Algorithms}

We have called the proposed algorithms greedy since they optimize a local objective function in each execution step~\cite{West2001}, namely the expected number of discoveries. In general, a greedy approach does not have to lead to optimal, or even good performance w.r.t.\@ any global performance goals. However, the proposed algorithms do achieve global optimality w.r.t.\@ several important objective functions. In particular, for \ac{BP} sets from $\mathbb{F}_3$ they generate recursive listening schedules and are thus complete, \ac{WDT}-optimal, \ac{MDT}-optimal, and \ac{NDoT}-optimal. For \ac{BP} sets from $\mathbb{F}_2$, they are still complete and \ac{WDT}-optimal. (They also achieve close-to-optimal performance w.r.t.\@ the \ac{MDT}, see Section~\ref{sec:num_exp}). These optimality results are proven in Section~\app{\ref{sec:optim_greedy} \SupplOf{}}\noapp{5 in~\cite{Karowski18_tech_report}}. Moreover, \ALG{} algorithms have a polynomial computational complexity (see Section~\app{\ref{sec:complexity} \SupplOf{}}\noapp{10 in~\cite{Karowski18_tech_report}} for details).

\subsection{Performance of CHAN TRAIN}

For \ac{BP} sets from $\mathbb{F}_3$, CHAN TRAIN is a \ALG{} algorithm and, consequently, inherits all the features of this family of algorithms. For \ac{BP} sets from $\mathbb{F}_2$, CHAN TRAIN is no longer \ALG{}. However, we are able to prove that it still achieves completeness and \ac{WDT}-optimality. These results are presented in Section~\app{\ref{sec:optim_chantrain} \SupplOf{}}\noapp{6 in~\cite{Karowski18_tech_report}}.

\subsection{Performance of \MDTOPT{}}

\MDTOPT{} is not only \ac{MDT}-optimal for arbitrary \ac{BP} sets but also \ac{WDT}-optimal for \ac{BP} sets from $\mathbb{F}_2$, since any \ac{MDT}-optimal schedule over $\mathbb{F}_2$ is also \ac{WDT}-optimal (see Section~\app{\ref{app:mdt_opt_cdt_opt} \SupplOf{}}\noapp{9 in~\cite{Karowski18_tech_report}} for a proof). Moreover, \MDTOPT{} is \ac{NDoT}-optimal for \ac{BP} sets from $\mathbb{F}_3$ (see Section~\app{\ref{app:mdt_opt_ndot_opt} \SupplOf{}}\noapp{9 in~\cite{Karowski18_tech_report}} for a proof). We remark, however, that \MDTOPT{} has a high computational complexity and memory consumption, and should only be performed offline and for network environments of moderate size. 

\subsection{Performance of \OPTBTwo{}}

\OPTBTwo{} generates recursive schedules for arbitrary \ac{BP} sets with two elements, and arbitrary numbers of channels. Consequently, it is complete, \ac{WDT}-optimal, \ac{MDT}-optimal, and \ac{NDoT}-optimal for any \ac{BP} set $B\in\mathbb{F}_1$ with $\lvert B\rvert=2$. This result is proven in Section~\app{\ref{sec:optim_optbtwo} \SupplIn{}}\noapp{7 in~\cite{Karowski18_tech_report}}.

\subsection{Performance of PSV}

\begin{figure}[t]
\centering
		\subfloat[PSV]{		
        \includegraphics[width=0.4\textwidth]{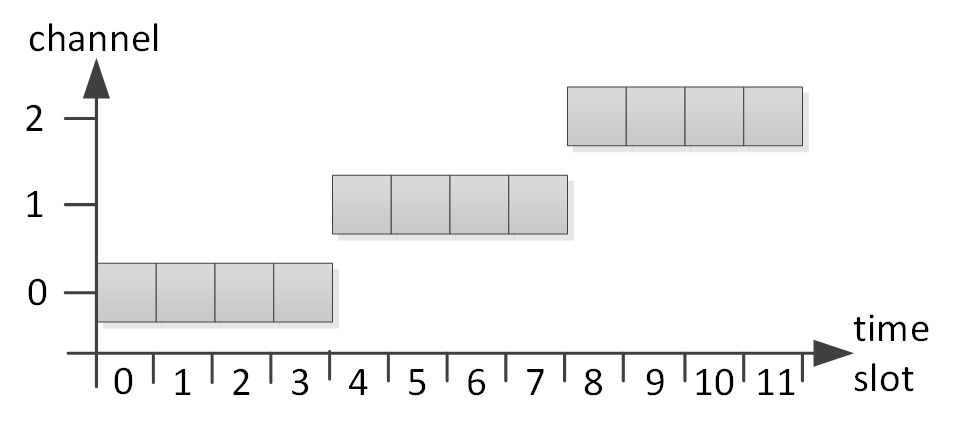}
				\label{fig:PSV_vs_GREEDY_MDT-PSV}
		} \\
		\subfloat[\GreedyRnd{}]{
        \includegraphics[width=0.4\textwidth]{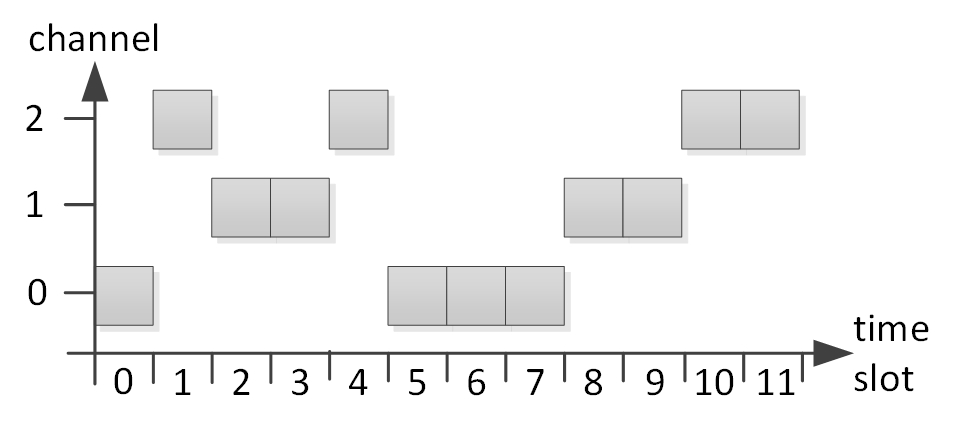}
				\label{fig:PSV_vs_GREEDY_MDT-GREEDYRND}
    }
\caption{Example with $B = \{1,2,4\} \in \mathbb{F}_3$ and $|C| = 3$; \GreedyRnd{} achieves optimal \ac{MDT}, while \ac{PSV} does not.}
\label{fig:GREEDY_vs_PSV_example}
\end{figure}

PSV is complete, \ac{WDT}-optimal and minimizes the number of channel switches for arbitrary \ac{BP} sets. However, it fails to optimize the \ac{NDoT} and \ac{MDT}, as shown by the example in Figure~\ref{fig:GREEDY_vs_PSV_example}, in which a uniform distribution of \acp{BP} is assumed. With $B=\{1,2,4\}$ and $|C|=3$, the schedule generated by PSV has a \ac{MDT} of $4.\overline{6}$, while \GreedyRnd{} achieves the optimal \ac{MDT} of $3$. Note that this example is very small in size -- the optimality gap may become arbitrarily large for scenarios with a higher number of channels and/or larger \ac{BP} sets.

\subsection{Summary}
\label{sec:strat_summ}

Table~\ref{tab:strategy_overview} summarizes the optimality and complexity results for the approaches presented in this section. 
For comparison, the table also contains discovery strategies OPT and SWOPT from our previous work~\cite{Karowski11, Karowski13}, as well as \ac{PSV}. 
 \section{Numerical Performance Evaluation}
\label{sec:num_exp}

We have performed numerical experiments to study settings and performance metrics not covered by the analytical results presented so far.

\subsection{Setting}
\label{sec:numexp_setting}

In order to evaluate the proposed algorithms over \ac{BP} sets from $\mathbb{F}_1$, $\mathbb{F}_2$, and $\mathbb{F}_3$ we draw random samples from these families. To include \MDTOPT{} in the evaluation of \ac{BP} sets  $\mathbb{F}_1$ and $\mathbb{F}_2$, we have to restrict the size of the studied scenarios, that is, the number of channels, as well as the number and magnitude of elements in the \ac{BP} sets. The latter is particularly important for sets from $\mathbb{F}_1$, since the complexity of \MDTOPT{} grows with $LCM(B)$ which is $\mathcal{O}\left(\prod_{b\in B}b\right)$.

We draw random samples $B\in\mathbb{F}_1$ as follows. We first draw the size $|B|$ of the \ac{BP} set from a uniform distribution over $\{3,\ldots,6\}$. We then draw individual \acp{BP} from a uniform distribution over $\{1,\ldots,10\}$. The selected \acp{BP} are then divided by their GCD (see Section~\ref{sec:BIFamilies}). The total number of \ac{BP} sets that can be obtained by this procedure is 775.

For $B\in\mathbb{F}_2$ we have to proceed differently in order to ensure the defining characteristic $\max(B)=LCM(B)$. For each number from $\{1,\ldots,256\}$ we first compute the power set of its factors. We then select subsets whose cardinality is uniformly distributed between 3 and 8, which contain the number itself, and whose GCD is one. We obtain 259286 sets.

For $B\in\mathbb{F}_3$ we draw samples as follows. First, we draw the size of $B$ from a uniform distribution over $\{2,\ldots,6\}$. Due to the definition of $\mathbb{F}_3$ and since we only consider \ac{BP} sets with \ac{GCD} of 1, we have $b_0=1$.
Then the other \acp{BP} are computed as $b_i=\prod_{j=1}^{i} x_j$, $i\in\{1,\ldots,|B|-1\}$, where $x_j$ are drawn from the uniform distribution over $\{2,\ldots,16\}$.
Due to the fact that \ALG{} algorithm are MDT-optimal for $B \in \mathbb{F}_3$, and in order to allow the evaluation of larger \acp{BP}, we have excluded \MDTOPT{} from experiments with \ac{BP} sets from $\mathbb{F}_3$.

Please note that due to the different approaches to randomly sample the corresponding family of \ac{BP} sets, the comparability of the results across the families of \ac{BP} sets is limited.

In addition to uniformly distributed channels and offsets, as described in Section~\ref{sec:system}, we assume a uniform distribution of \acp{BP}. That is, the probability that a neighbor is using a certain \ac{BP} is $1/{|B|}$. Consequently, the probability of the configuration $\kappa$ to be selected by a neighbor is $P_\kappa=\frac{1}{b_\kappa|B||C|}$.

We vary the number of channels between 2 and 12. For each number of channels, we perform at least 150 runs with randomly selected \ac{BP} sets.

\subsection{Results}
\label{sec:numerical_ev_results}

\begin{figure*}
\centering
\renewcommand{\thesubfigure}{a}\subfloat[\acf{MDT} $\mathbb{F}_3$]{
\includegraphics[width=\evalFigWidth\textwidth]{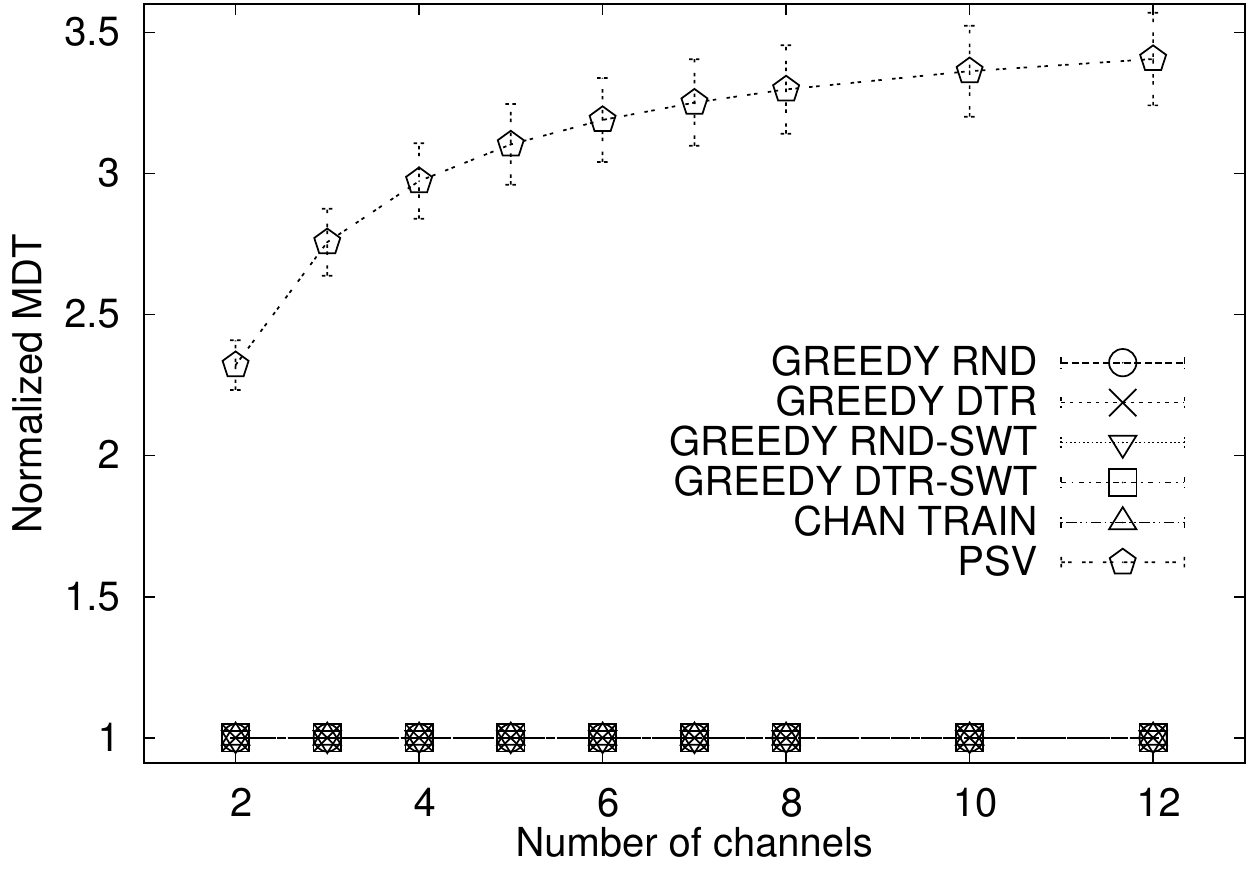}
\label{fig:F3_discTime}
}\renewcommand{\thesubfigure}{b}\subfloat[\acf{MDT} $\mathbb{F}_2$]{
\includegraphics[width=\evalFigWidth\textwidth]{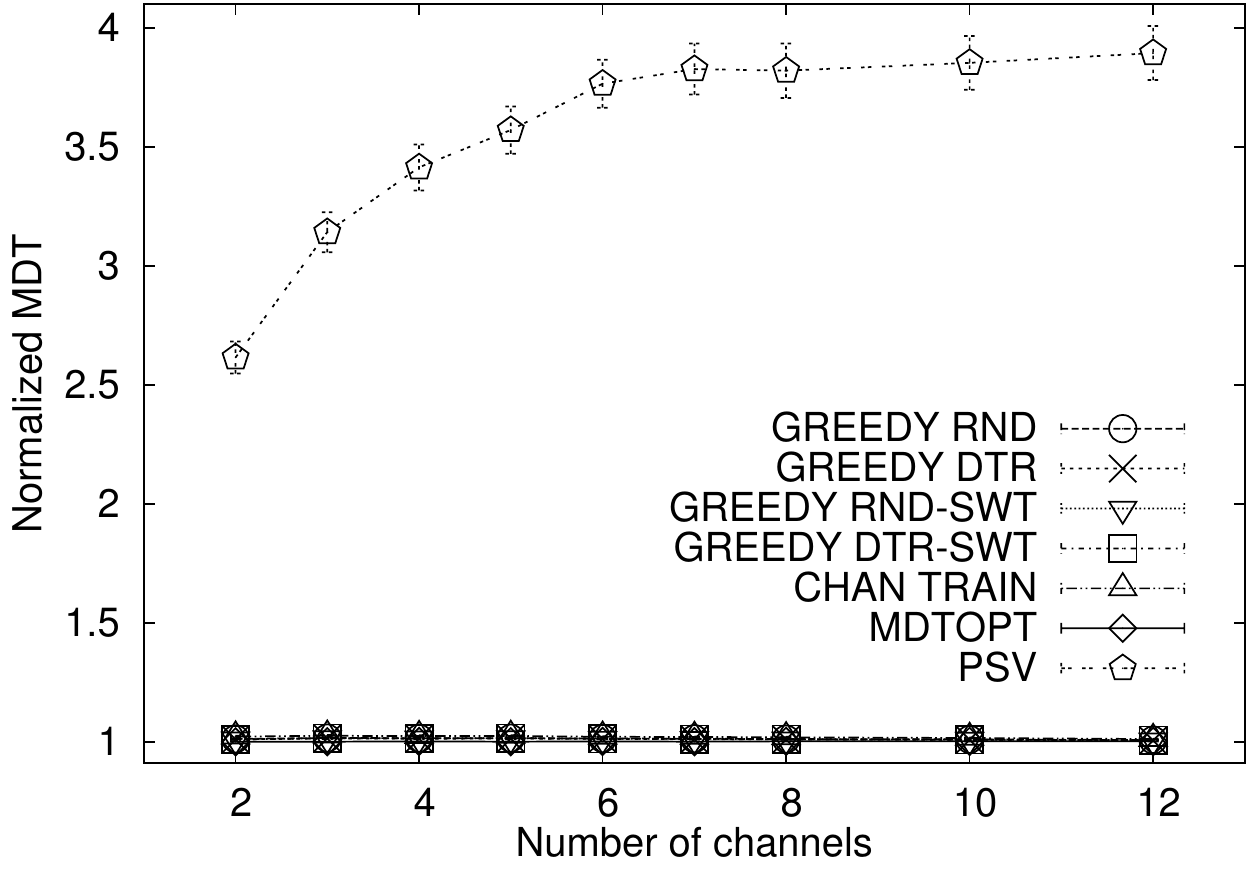}
\label{fig:F2_discTime}
}\renewcommand{\thesubfigure}{c}\subfloat[\acf{MDT} $\mathbb{F}_1$]{
\includegraphics[width=\evalFigWidth\textwidth]{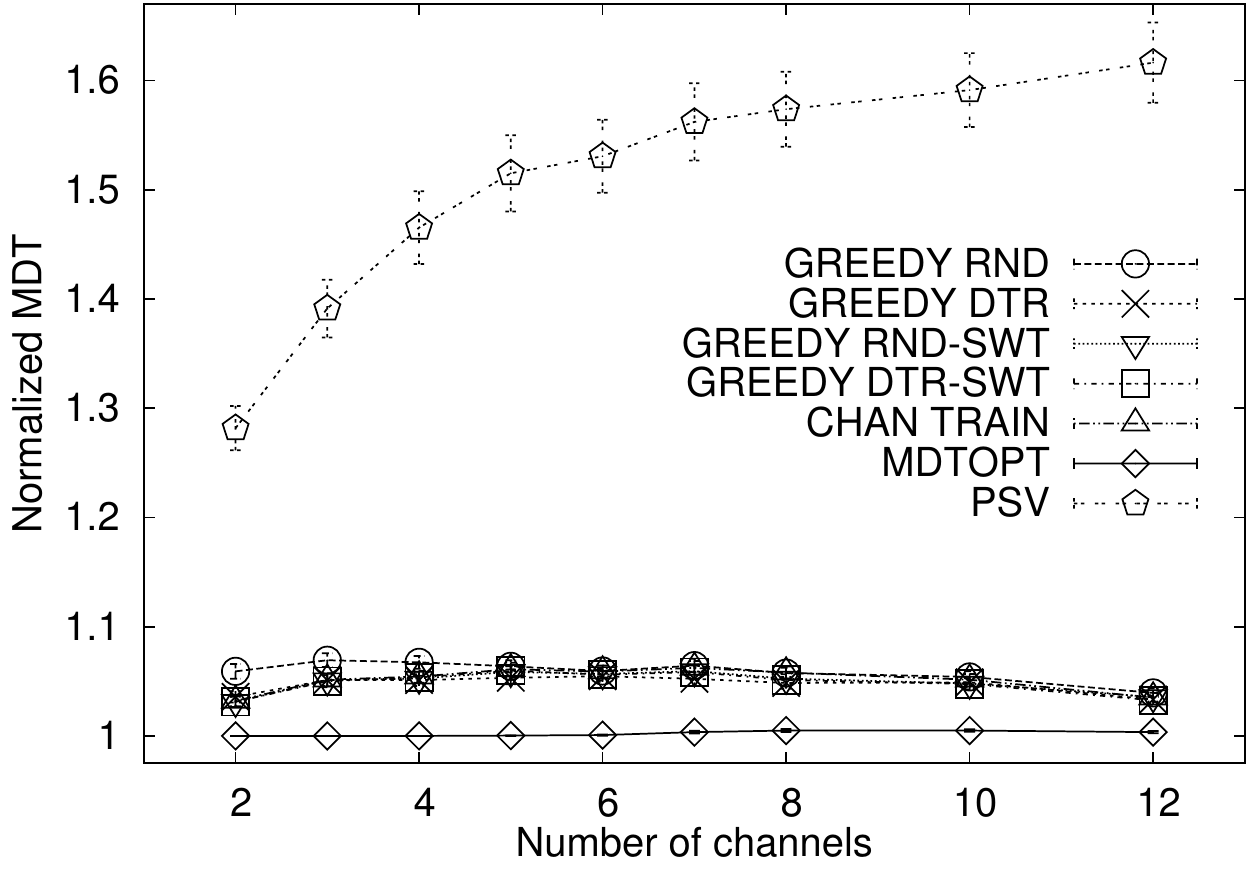}
\label{fig:F1_discTime}
}\\
\renewcommand{\thesubfigure}{d}\subfloat[\acf{NDoT} $\mathbb{F}_3$]{		
\includegraphics[width=\evalFigWidth\textwidth]{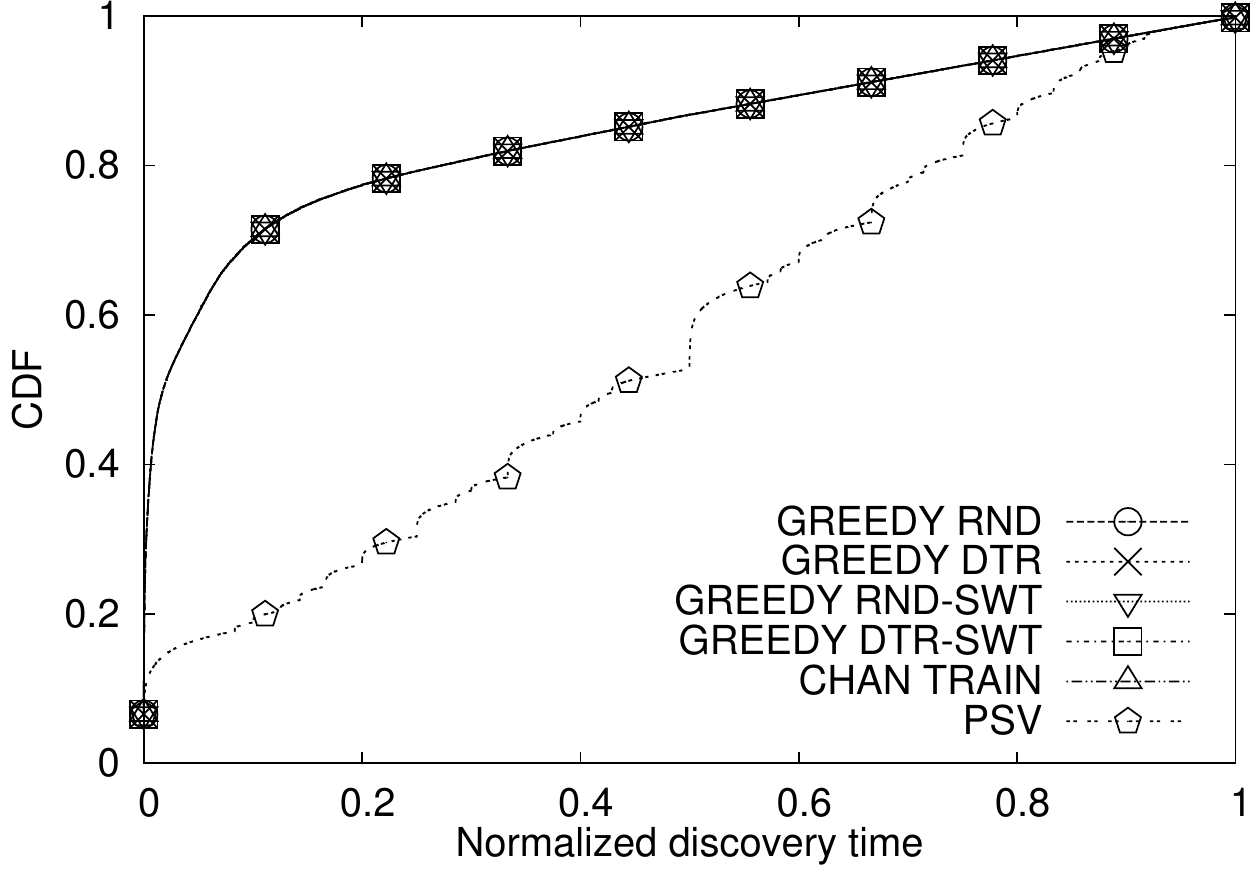}
\label{fig:F3_MDOT_numerical}
}
\renewcommand{\thesubfigure}{e}\subfloat[\acf{NDoT} $\mathbb{F}_2$]{	
\includegraphics[width=\evalFigWidth\textwidth]{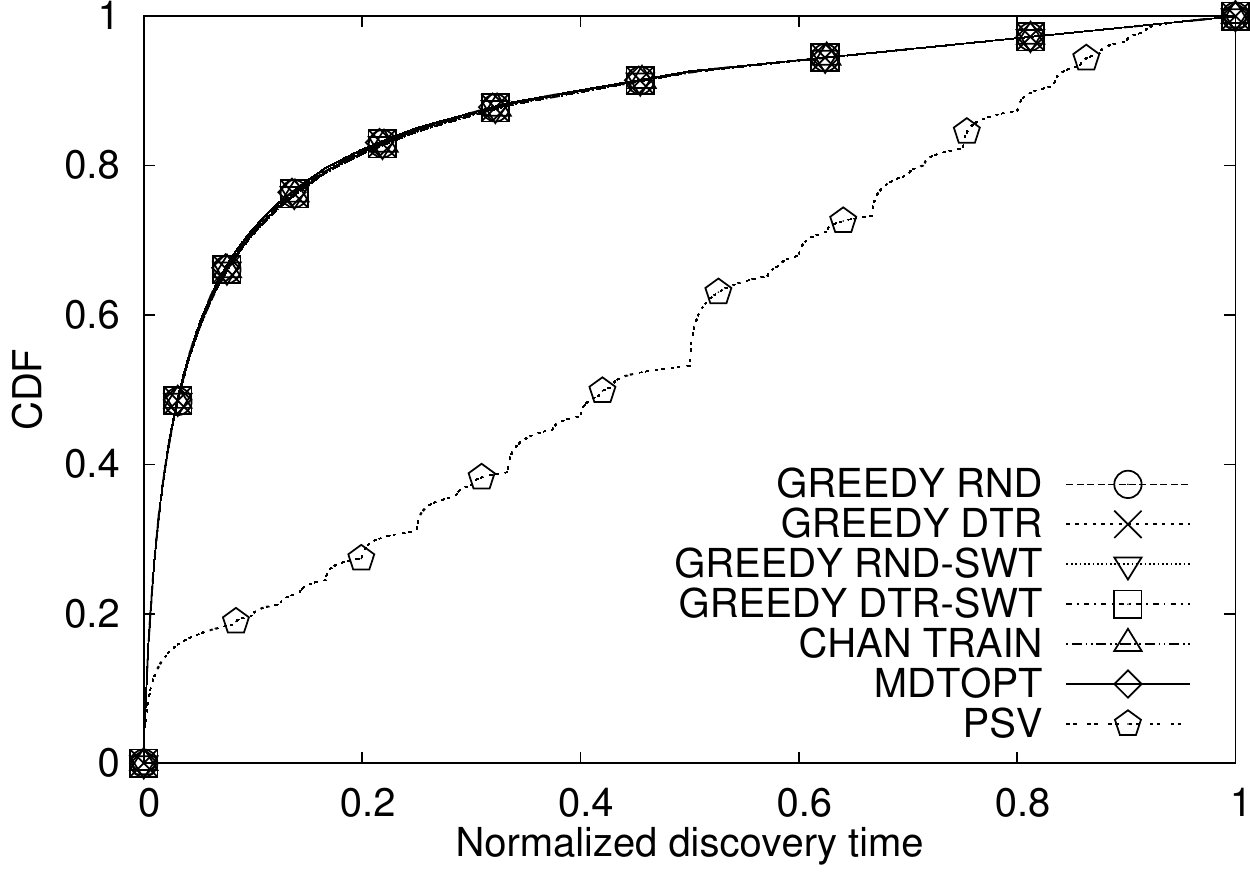}
\label{fig:F2_MDOT_numerical}
}\renewcommand{\thesubfigure}{f}\subfloat[\acf{NDoT} $\mathbb{F}_1$]{		
\includegraphics[width=\evalFigWidth\textwidth]{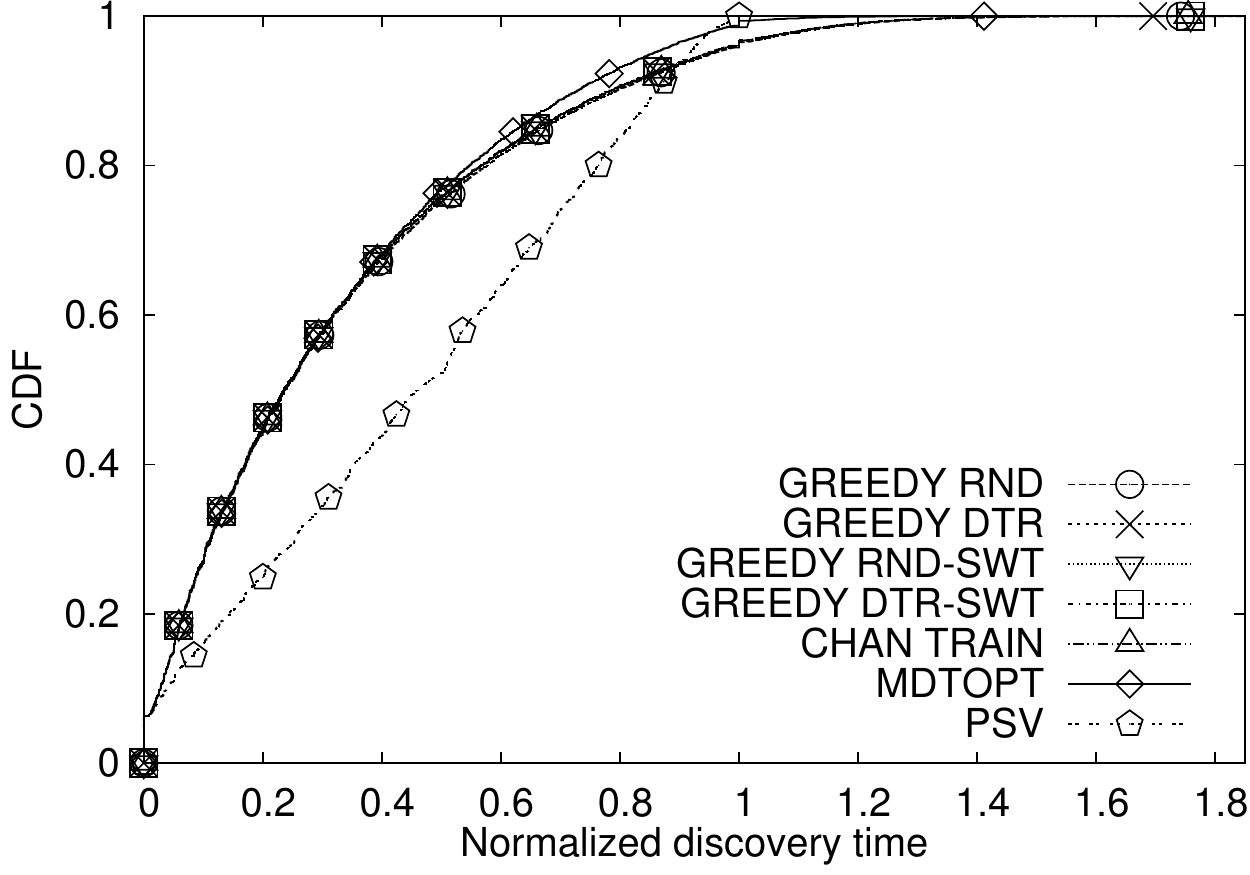}
\label{fig:F1_MDOT_numerical}
}\\
\renewcommand{\thesubfigure}{g}\subfloat[Number of Channel Switches $\mathbb{F}_3$]{		
\includegraphics[width=\evalFigWidth\textwidth]{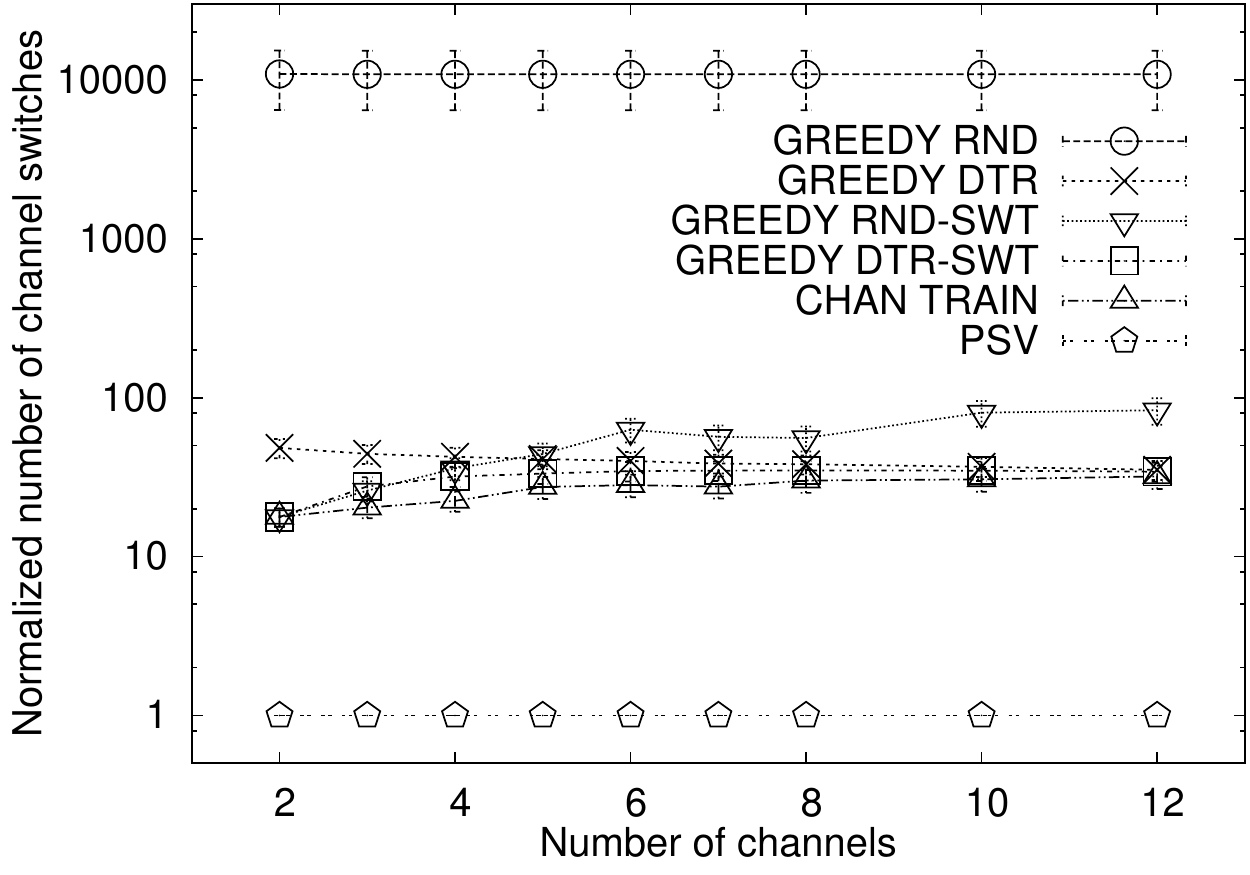}
\label{fig:F3_chanJumps}
}
\renewcommand{\thesubfigure}{h}\subfloat[Number of Channel Switches $\mathbb{F}_2$]{	
\includegraphics[width=\evalFigWidth\textwidth]{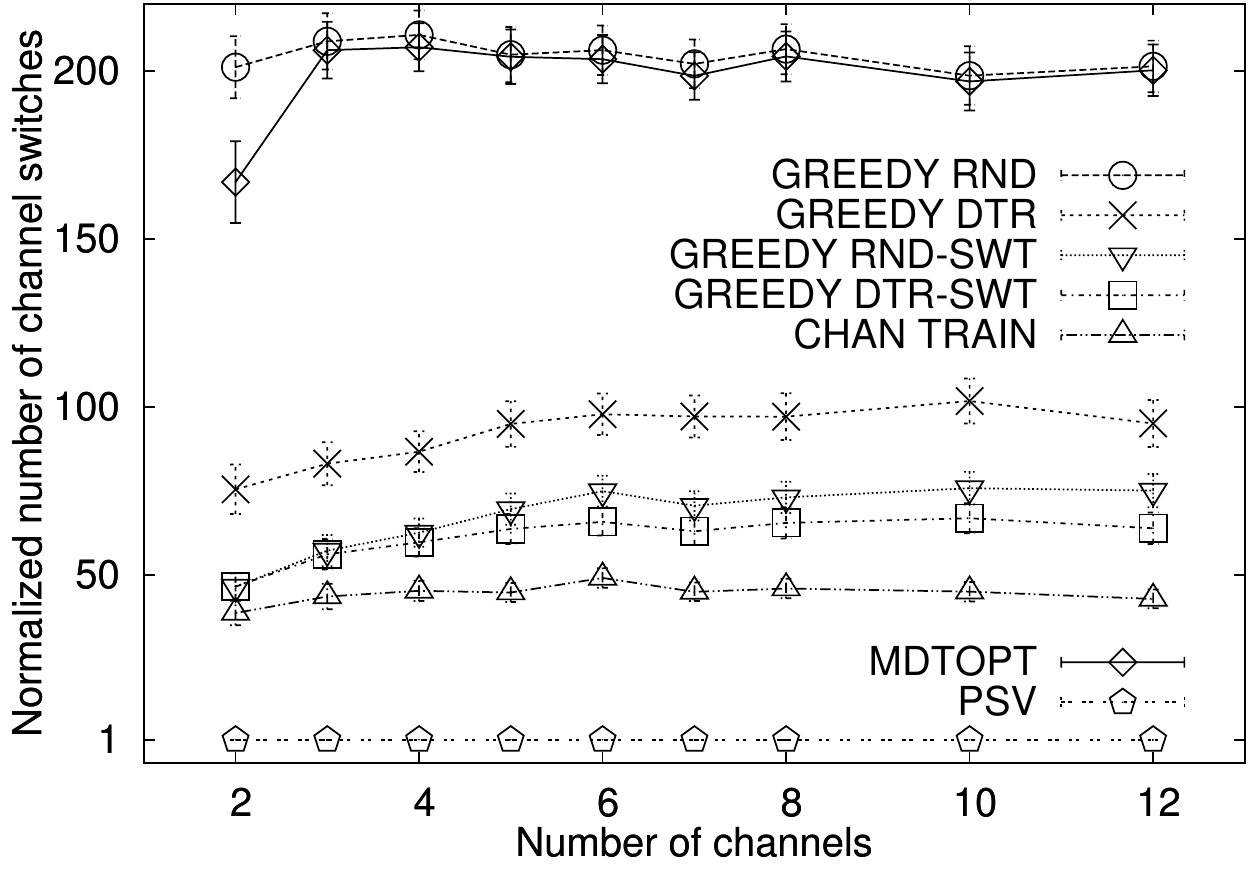}
\label{fig:F2_chanJumps}
}\renewcommand{\thesubfigure}{i}\subfloat[Number of Channel Switches $\mathbb{F}_1$]{		
\includegraphics[width=\evalFigWidth\textwidth]{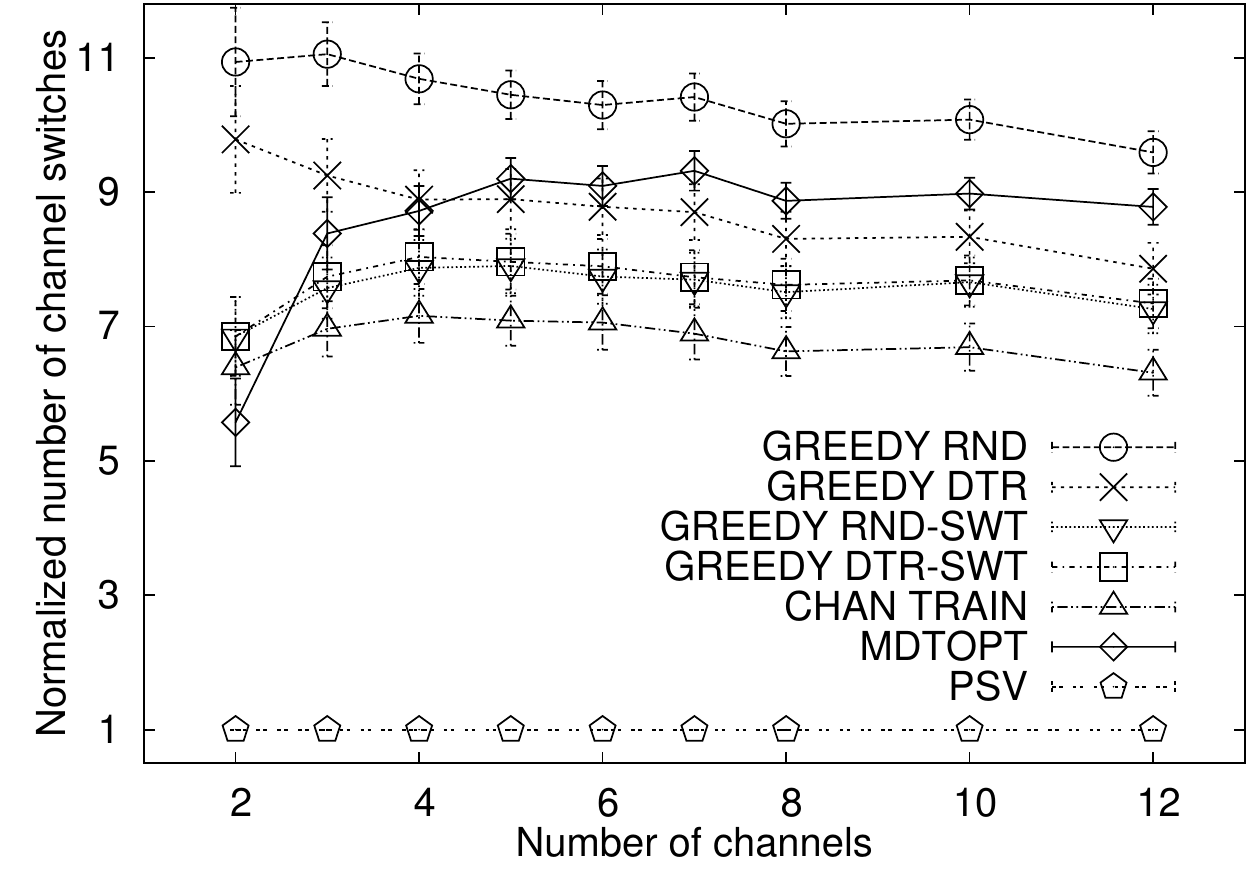}
\label{fig:F1_chanJumps}
}\caption{Numerical evaluation results of the normalized \ac{MDT}, normalized \ac{NDoT} and normalized number of channel switches (see Sections~\ref{subsec:num_results_mdt},~\ref{subsec:num_results_mdot} and ~\ref{subsec:num_results_chan_switches}).}
\label{fig:numeric_results_1}
\end{figure*}

In the following, in order to compare the results across different \ac{BP} sets, all results are normalized to the respective optimum values as described in the following sections. Further, all results are accompanied by confidence intervals for a confidence level of 95\%.

\subsubsection{Results for \ac{MDT}}
\label{subsec:num_results_mdt}

Figure~\ref{fig:F3_discTime} depicts the \ac{MDT} normalized to the optimum value obtained by executing any \ALG{} algorithm, for \ac{BP} sets from $\mathbb{F}_3$. As expected, all \ALG{} strategies and CHAN TRAIN achieve optimum \ac{MDT}. At the same time, even though PSV uses the same number of listening slots to perform a complete discovery, and has the same (optimal) \ac{WDT} as our approaches, it results in a \ac{MDT} which is by more than 300\% higher, while the gap is further increasing for a larger number of channels.

Figure~\ref{fig:F2_discTime} depicts the \ac{MDT} normalized to their optimum values as computed by \MDTOPT{}, for \ac{BP} sets from $\mathbb{F}_2$. We observe that \ALG{} algorithms are within 2\% of the optimum and approximate the optimum even further when the number of channels increases. The \ac{MDT} of CHAN TRAIN is still within 3\% of the optimum. In contrast, PSV has a significantly larger \ac{MDT}, reaching 400\% of the optimum, and diverging when the number of channels increases.

Figure~\ref{fig:F1_discTime} shows the normalized \ac{MDT} for the family of \ac{BP} sets $\mathbb{F}_1$. We observe that the performance of the individual discovery algorithms relative to each other does not significantly differ from the $\mathbb{F}_2$ case. We also observe that their normalized \ac{MDT} has slightly increased. A potential explanation for this is, on the one hand, the less regular structure of \ac{BP} sets in $\mathbb{F}_1$.
On the other hand, the evaluation for \ac{BP} sets from $\mathbb{F}_1$ is performed with considerably smaller \acp{BP}, in order to make the computation of \MDTOPT{} feasible. Still, while the \ac{MDT} of the \ALG{} strategies are within 7\% of the optimum, the \ac{MDT} of PSV reaches 160\% of the optimum and is increasing with the number of channels.

\subsubsection{Results for \ac{NDoT}}
\label{subsec:num_results_mdot}

Figures~\ref{fig:F3_MDOT_numerical} and~\ref{fig:F2_MDOT_numerical} show the \acp{CDF} of the normalized discovery times for \ac{BP} sets of from $\mathbb{F}_3$ and $\mathbb{F}_2$, for 2 to 12 channels. Due to the more frequent channel switching the \ALG{} approaches need less time to discover more the individual neighbors. For example, after the first 10\% to 20\% of the schedule is executed (normalized discovery time between 0.1 and 0.2), \ALG{} discovers up to 50\% more neighbors than PSV on the average.
CHAN TRAIN and \MDTOPT{} achieve an equivalent performance as \alg{}.

\begin{figure}[t]
\centering
\renewcommand{\thesubfigure}{a}\subfloat[\acf{WDT} $\mathbb{F}_1$]{
\includegraphics[width=\evalFigWidth\textwidth]{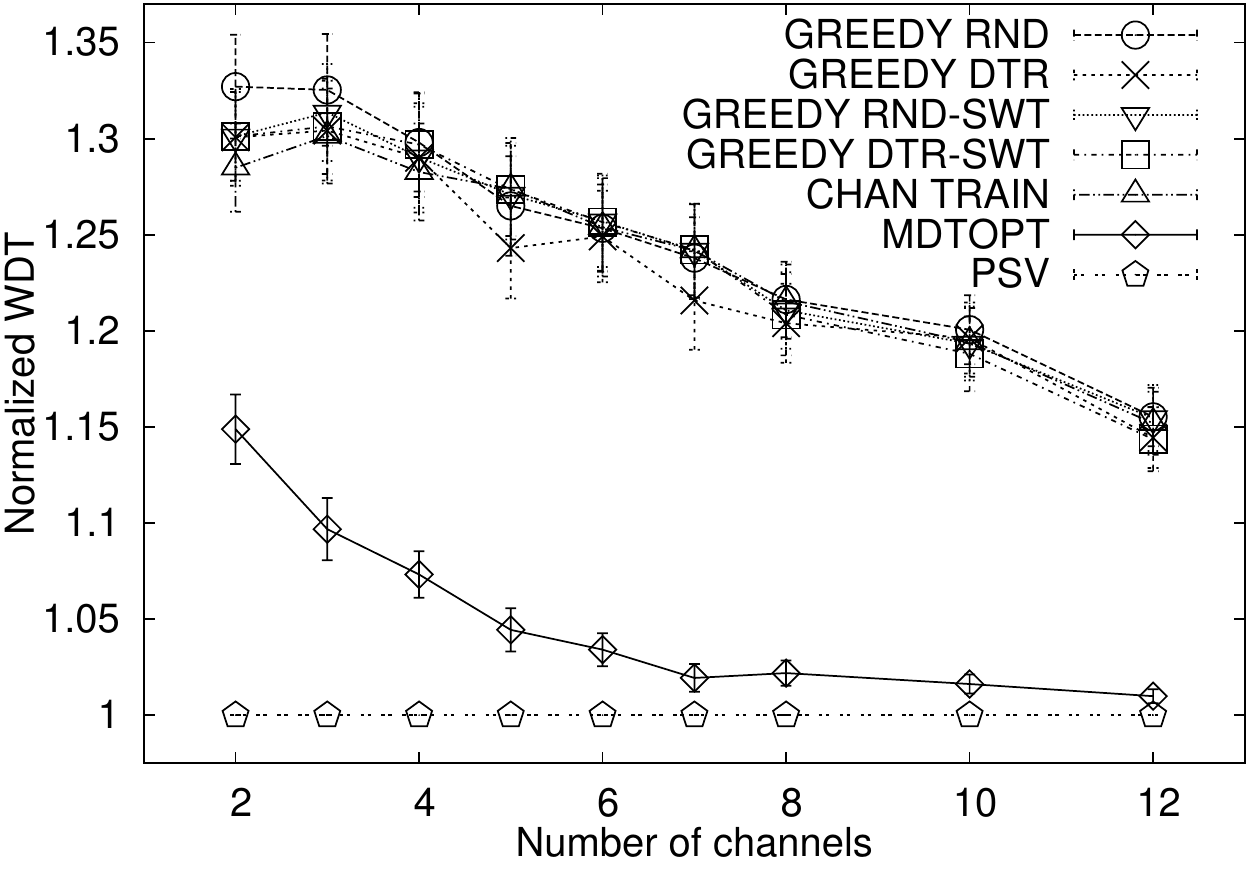}
\label{fig:F1_makeSpan}
}
\\
\renewcommand{\thesubfigure}{b}\subfloat[Number of Listening Time Slots $\mathbb{F}_1$]{
\includegraphics[width=\evalFigWidth\textwidth]{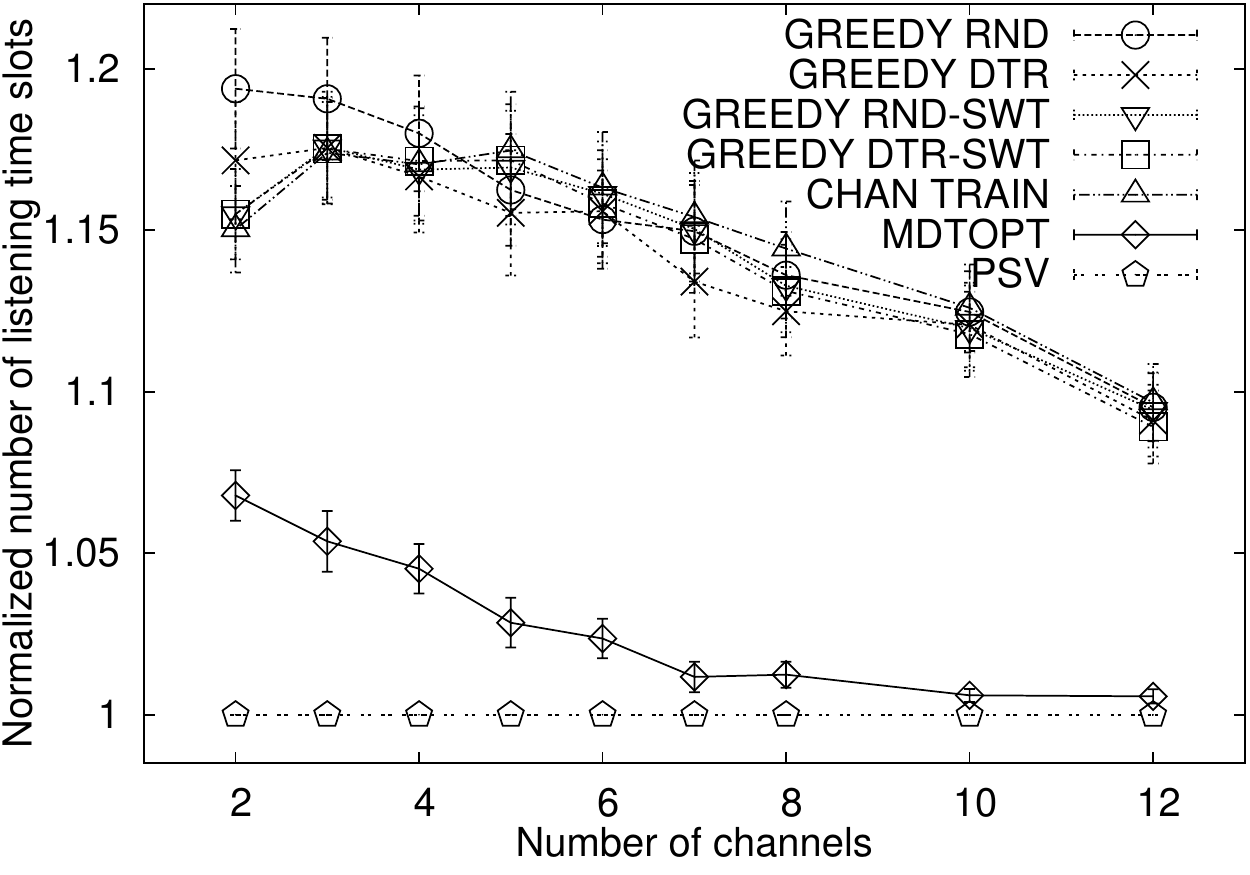}
\label{fig:F1_listenSlots}
}	
\caption{Numerical evaluation results of the normalized \ac{WDT} and normalized number of listening time slots for the families of \ac{BP} set $\mathbb{F}_1$ (see Section~\ref{subsec:num_results_cdt_active_slots}).}
\label{fig:numeric_results_2}
\end{figure}

Figure~\ref{fig:F1_MDOT_numerical} displays the results for \ac{BP} sets from $\mathbb{F}_1$. Even for this most general family of \ac{BP} sets the presented approaches discover by up to 20\% more neighbors until certain points in time, as compared to PSV. However, the discovery of the last 10\% of neighbors takes more time than with \ac{PSV}, which is also reflected in Figure~\ref{fig:F1_makeSpan} depicting the performance w.r.t. the \ac{WDT} (see Section~\ref{subsec:num_results_cdt_active_slots} for details).

\subsubsection{Results for number of channel switches}
\label{subsec:num_results_chan_switches}

Figures~\ref{fig:F3_chanJumps},~\ref{fig:F2_chanJumps}, and~\ref{fig:F1_chanJumps} depict the number of channel switches normalized to the minimum value $\lvert C\rvert -1$ for the families of \ac{BP} sets $\mathbb{F}_3$, $\mathbb{F}_2$, and $\mathbb{F}_1$. We have include this metric in the evaluation since a high number of switches may reduce the success rate if the deaf periods during the switches are long enough. 

Due to its random selection of channels among the candidates, \GreedyRnd{} results in the highest number of channel switches.
At the same time, the design of CHAN TRAIN aiming at reducing the number of channel switches is successful in achieving its goal. It results in the second lowest values.
We can also observe that a simple prioritization of the channel allocated during the previous time slot, as performed by \GreedyTrainRnd{} and \GreedyTrainDeter{}, allows to significantly reduce the number of channel switches.

We remark that even though a high number of channel switches has the potential to decrease the success rate in realistic scenarios, our simulations reveal that simple mechanisms that reduce channel switches such as \GreedyTrainRnd{} and \GreedyTrainDeter{} already achieve an optimal success rate (see Section~\ref{sec:simus} for details).

\subsubsection{Results for \ac{WDT} and number of listening time slots}
\label{subsec:num_results_cdt_active_slots}

Since all considered strategies achieve an optimal \ac{WDT} and therefore also an optimal number of listening time slots for the families of \ac{BP} sets $\mathbb{F}_2$ and $\mathbb{F}_3$ (see Section~\ref{sec:perf_anal} for details), we are evaluating this metric only for \ac{BP} sets from $\mathbb{F}_1$.

Figure~\ref{fig:F1_makeSpan} shows the \ac{WDT}, normalized to its optimum value $\max(B)\lvert C\rvert$. For all strategies it improves with the increasing number of channels. For \MDTOPT{} the gap reduces from about 15\% for two channels to about 1\% for 12 channels. Note that despite the upper bound of $LCM(B)|C|$ time slots required to optimize \ac{MDT}, on average it takes only 15\% (respective 1\%) more time slots than the optimum value of $\max(B)|C|$.
Listening schedules generated by \ALG{} and CHAN TRAIN result in similar \ac{WDT}.

Since the \ac{WDT} of a schedule can be significantly larger than its actual energy usage due to idle slots, Figure~\ref{fig:F1_listenSlots} depicts the number of listening time slots normalized to its optimum value $\max(B)\lvert C\rvert$. The results for all strategies are very similar to those for the \ac{WDT}, except that they are shifted by about 5\% meaning that the schedules computed by \ALG{} approaches, CHAN TRAIN, and \MDTOPT{}, consist of about 5\% idle slots in which no scan is scheduled on any channel due to the fact that no new configurations can be discovered.

We remark that out of the three main considered performance metrics the \ac{WDT} over $\mathbb{F}_1$ is the only performance metric and \ac{BP} family for which \ac{PSV} offers better performance than the proposed \alg{} discovery approaches. In addition, the performance gap decreases for higher numbers of channels.

\subsection{Summary}

From the results of the numerical experiments we observe that w.r.t. the \ac{MDT} the \ALG{} algorithms and CHAN TRAIN significantly (by up to several hundreds percent) outperform PSV for all families of \ac{BP} sets. Furthermore, in addition to being optimal for \ac{BP} sets from $\mathbb{F}_3$, they achieve close-to-optimal \ac{MDT} for \ac{BP} sets from $\mathbb{F}_2$ and even $\mathbb{F}_1$. Also w.r.t. the \ac{NDoT}, the proposed algorithms significantly outperform \ac{PSV} for all families of \ac{BP} sets.

Furthermore, we observe that the approaches aiming at heuristically reducing the number of channel switches succeed in achieving this goal (except for \GreedyTrainRnd{} in  $\mathbb{F}_3$), even though they still require considerably more switches than the optimum value. We remark, however, that the large number of channel switches does not necessarily have a negative impact on the performance (see also Section~\ref{sec:simus}).

Finally, we observe that the proposed approaches are all within 30\% of the optimum for the \ac{WDT} over $\mathbb{F}_1$, while this gap further decreases for higher numbers of channels. The number of listening time slots, which is proportional to the energy consumption, is within 20\% of the optimum for all considered approaches, while the gap, again, decreases for larger numbers of channels. 
We remark that all studied approaches have optimal \ac{WDT} for \ac{BP} families $\mathbb{F}_2$ and $\mathbb{F}_3$.

\section{Evaluation under Realistic Conditions}
\label{sec:simus}

For the design and analysis of the proposed discovery algorithms, we have made idealizing assumptions such as the absence of beacon losses due to collisions, interference, and deaf periods caused by channels switches, and zero beacon transmission/reception times. In reality, these assumptions do not hold which may reduce the success rate, which is the fraction of discovered neighbors after the first execution of the listening schedule. Consequently, we perform a series of simulations under realistic conditions, described in the following, in order to evaluate the success rate.
In addition, Section~\app{\ref{sec:appendix_results} \SupplOf{}}\noapp{11 in~\cite{Karowski18_tech_report}} contains further simulation results.

\subsection{Setting}
\label{subsec:sim_eval_settings}

We use the OMNeT++ 3.3 simulator~\cite{omnet} together with the Mobility Framework (MF)~\cite{mf-omnet} and a model of the IEEE 802.15.4 PHY and MAC layers, extended to support arbitrary \acp{BP}. The wireless model and a description of the implementation are provided in~\cite{Karowski13}.

In contrast to the settings used to achieve the analytical and numerical results presented in the previous sections, in simulations we drop the idealizing assumptions described in Section~\ref{sec:system}. The PHY frame size of beacons used in the simulations is 19 bytes, corresponding to an empty MAC payload. The channel switching time is set to 24 symbols matching the settling time between channel switches of a commonly used IEEE~802.15.4 radio chip~\cite{AT86-datasheet}.

We draw 250 random samples from $\mathbb{F}_1$ and $\mathbb{F}_2$, respectively, as described in Section~\ref{sec:numexp_setting}, except that we now study a broader range of \ac{BP} sets from $\mathbb{F}_1$. To be able to do that we exclude \MDTOPT{} from the evaluation due to its high computational complexity. 
Moreover, the maximum cardinality of the \ac{BP} sets is set to 6, while the maximum \ac{BP} is set to 128. This results in approximately $5.7\cdot 10^9$ candidate \ac{BP} sets for $\mathbb{F}_1$ and approximately $1.6\cdot 10^4$ for $\mathbb{F}_2$. While we significantly increase the size of considered sample from $\mathbb{F}_1$, we reduce the number of considered sets from $\mathbb{F}_2$, in order to make the results for these two families of \ac{BP} sets better comparable.

To study the dependence of the performance on the number of neighbors, we vary the number of neighbors between 2 and 35, randomly placing them within the communication range of the discoverer. The number of channels is hereby fixed at 8. To study the dependence on the number of channels, we vary the number of channels between 2 and 12, while keeping the number of neighbors fixed at 15.

Each neighbor $\nu\in N$ randomly selects a channel $c_\nu$ and a \ac{BP} $b_\nu$ from uniform distributions over $C$ and $B$. 
The start time $t_{\nu}$ of each neighbor $\nu$ is uniformly distributed over $[0; \tau b_{\nu}[$; the resulting offset is given by $\delta_{\nu} = \lfloor t_{\nu} / \tau \rfloor$. The discoverer starts operation at time slot 0 and executes the listening schedule once. For each \ac{BP} set, we perform 5 runs with randomly selected configurations such that each statistic in the following is computed over 1250 runs (250 sets, 5 runs per set). 

\subsection{Results}
\label{subsec:sim_results}

\begin{figure}
\centering
\subfloat[Success Rate $\mathbb{F}_2$]{
\includegraphics[width=\evalFigWidth\textwidth]{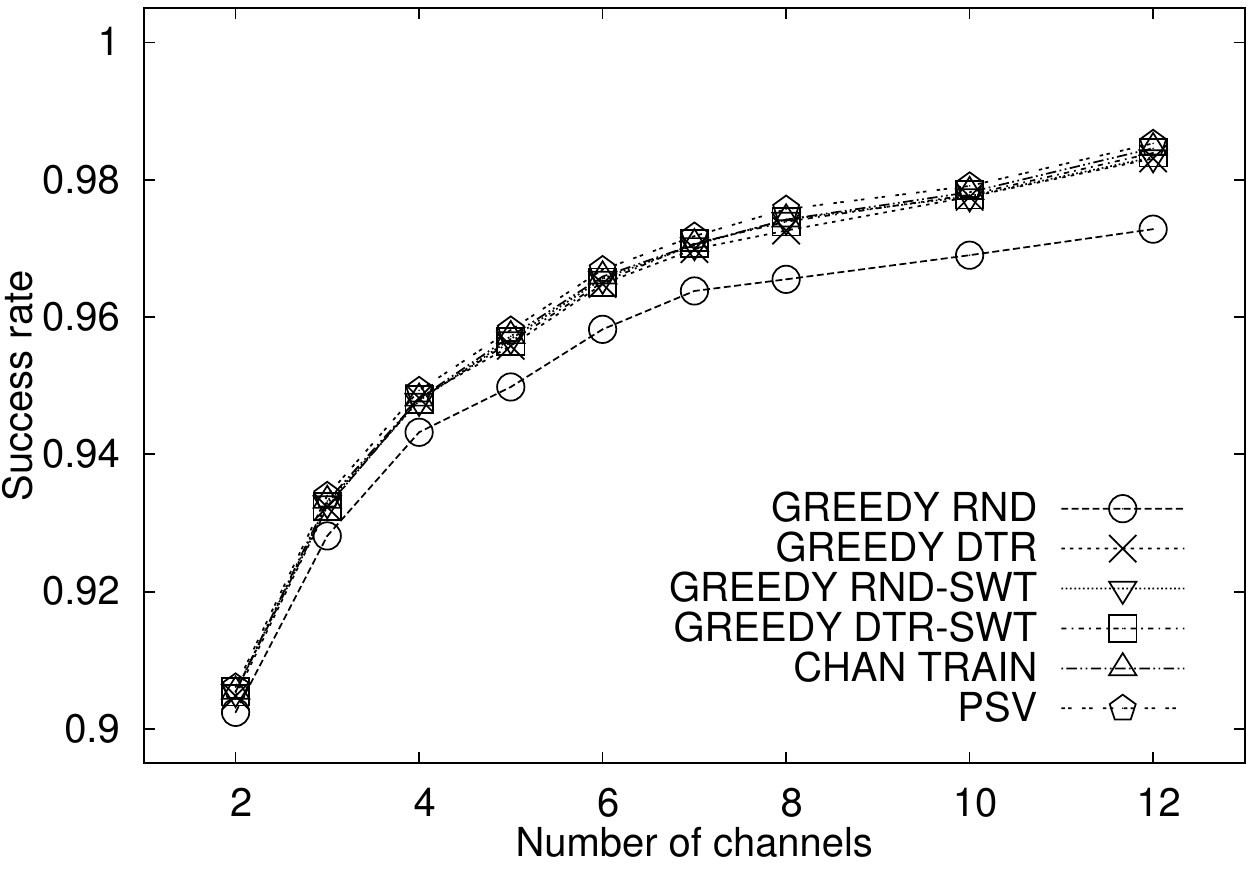}
\label{fig:F2_numChan_discFreq_sim}
}
\\
\subfloat[Success Rate $\mathbb{F}_2$]{
\includegraphics[width=\evalFigWidth\textwidth]{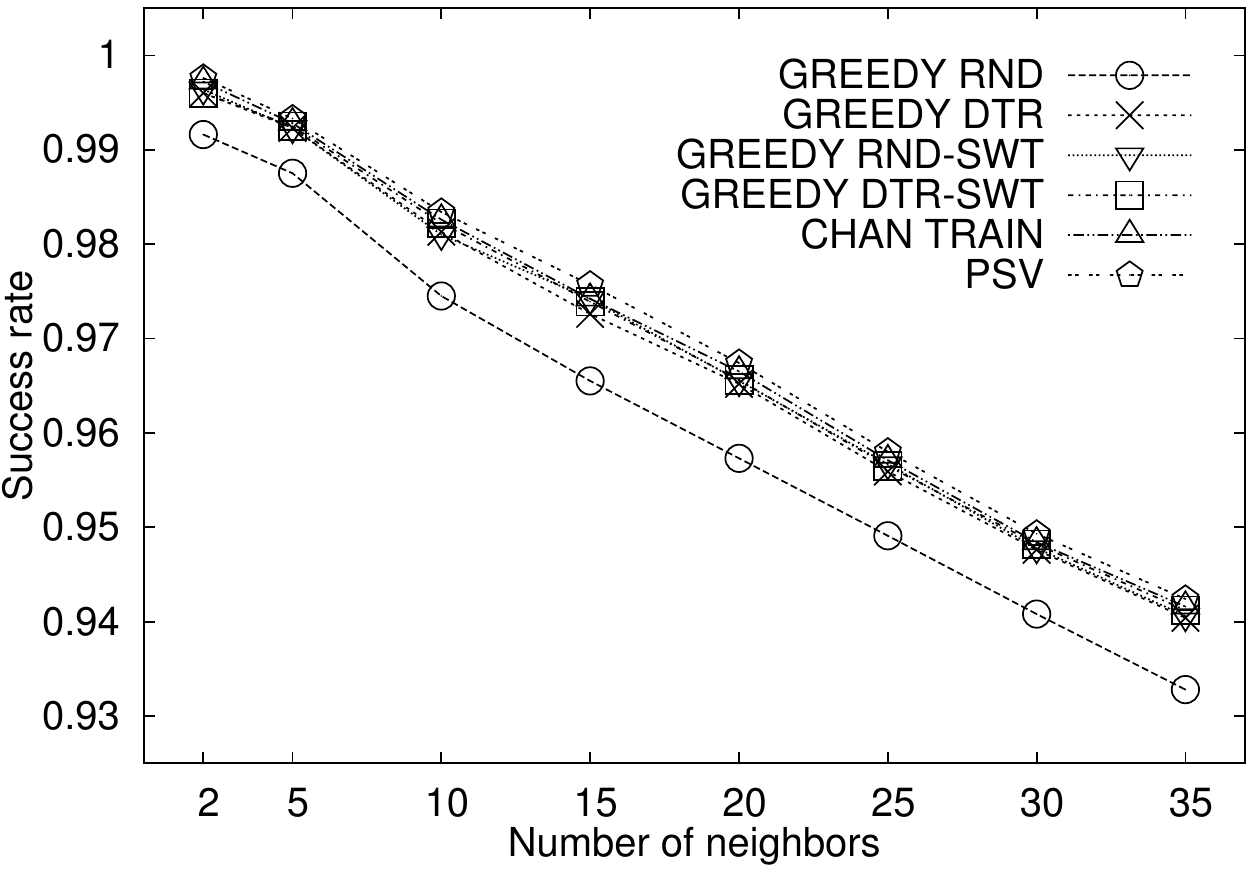}
\label{fig:F2_numConf_discFreq_sim}
}
\caption{Evaluation results of the success rate obtained by simulation for the family of \ac{BP} sets $\mathbb{F}_2$ (see Section~\ref{subsec:sim_results}). }
\label{fig:F2_sim_results_success_rate}
\end{figure}

\begin{figure}
\centering
\subfloat[Success Rate $\mathbb{F}_1$]{
\includegraphics[width=\evalFigWidth\textwidth]{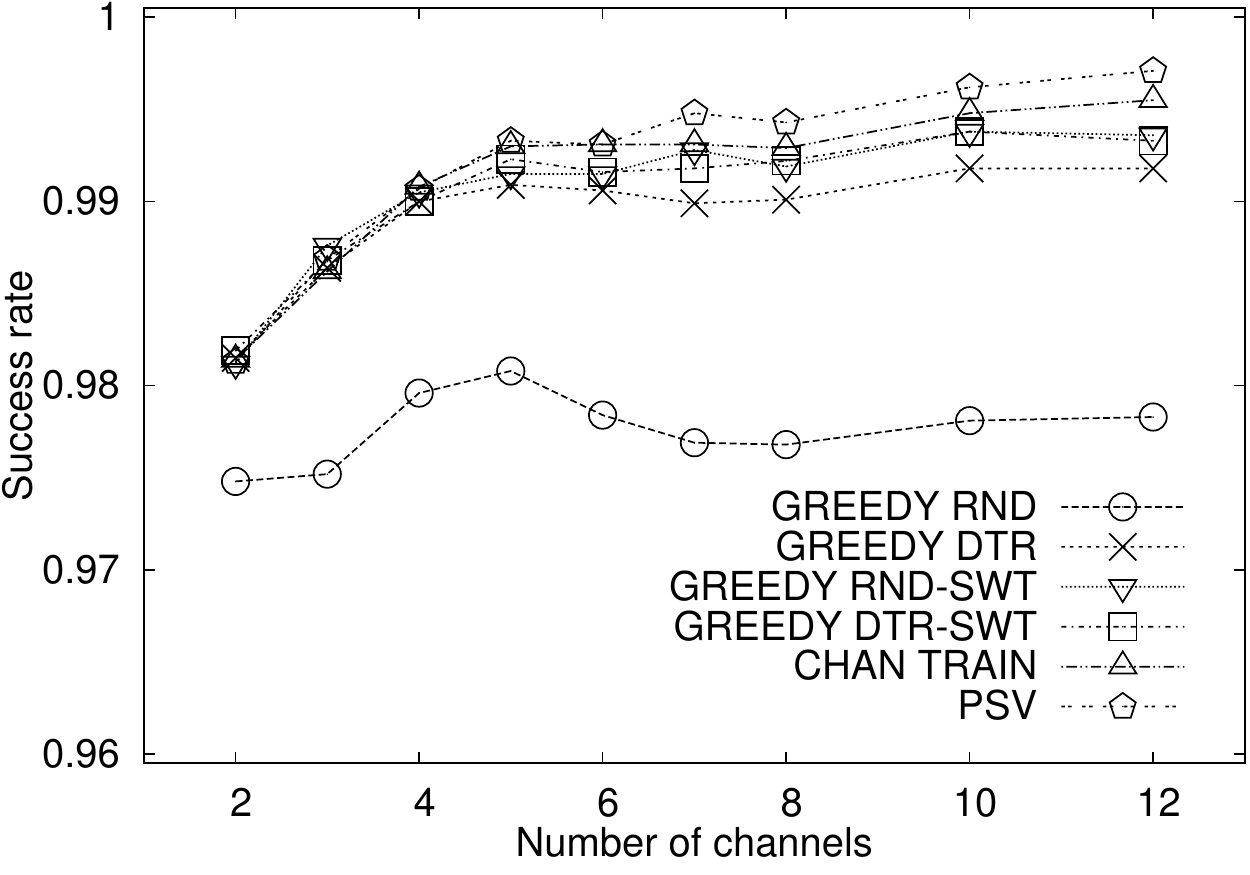}
\label{fig:F1_numChan_discFreq_sim}
}
\\
\subfloat[Success Rate $\mathbb{F}_1$]{
\includegraphics[width=\evalFigWidth\textwidth]{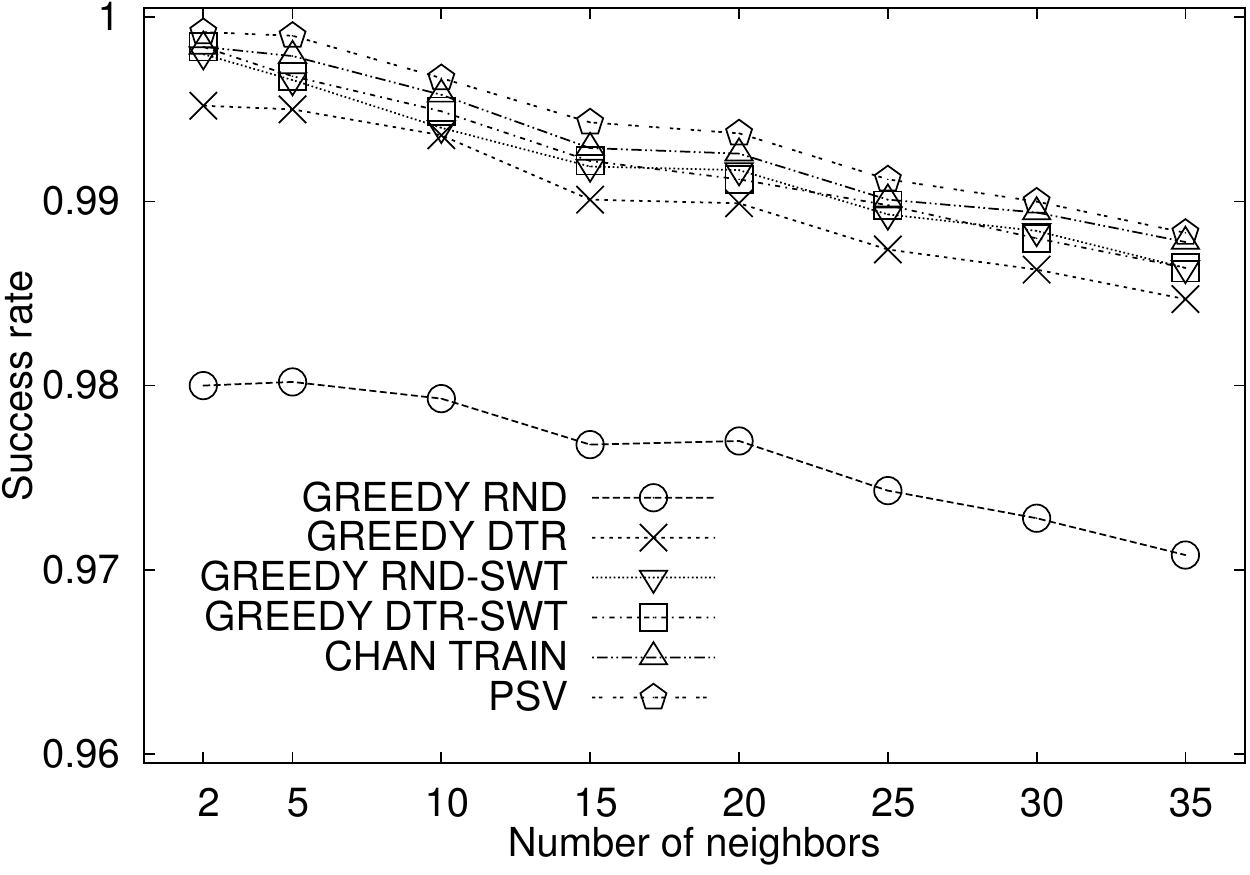}
\label{fig:F1_numConf_discFreq_sim}
}
\caption{Evaluation results of the success rate obtained by simulation for the family of \ac{BP} sets $\mathbb{F}_1$ (see Section~\ref{subsec:sim_results}).}
\label{fig:F1_sim_results_success_rate}
\end{figure}

The success rate for the family of \ac{BP} sets $\mathbb{F}_2$ and $\mathbb{F}_1$ is depicted in Figures~\ref{fig:F2_sim_results_success_rate} and~\ref{fig:F1_sim_results_success_rate}, respectively. With an increasing number of channels all strategies result in a significantly higher success rates due to the lower probability of overlapping beacon transmissions. A reverse effect is observed when the number of neighbors increases. Furthermore, over $\mathbb{F}_1$ all strategies achieve a higher success rate than over $\mathbb{F}_2$, which is caused by the reduced probability of colliding beacon transmissions due to the structure of \ac{BP} sets from the family $\mathbb{F}_1$. These effects are independent of the deployed discovery algorithm. 

Except for \GreedyRnd{}, the strategies result in similar values, while CHAN TRAIN is closest to the optimum represented by \ac{PSV}. The lower success rate of \GreedyRnd{} is caused by the high number of channel switches of this strategy, causing many deaf periods. Note that a decrease in success rate is the only potential negative impact of the higher number of channels switches. Thus, this results reveals that the potential drawback of a higher number of channel switches exhibited by the \ALG{} and \MDTOPT{} discovery strategies as compared to the PSV discovery strategy is negligible, if appropriate heuristics are performed, as done by \ALG{} RND-SWT, \ALG{} DTR-SWT, and CHAN TRAIN.

\subsection{Summary}

The evaluation under realistic conditions reveals that the impact of a higher number of channel switches performed by the developed approaches in order to minimize the discovery times can be successfully counteracted by simple heuristics such as the ones deployed by \GreedyTrainRnd{}, \GreedyTrainDeter{}, and CHAN TRAIN.
 \section{\aclp{BP} Supporting Efficient Neighbor Discovery}
\label{sec:biSelection}

As observed in the previous sections, there is a trade-off between the flexibility of the used \ac{BP} sets and the efficiency of the discovery process. It is possible to minimize the \ac{WDT} or the number of channel switches for arbitrary \ac{BP} sets $B\in\mathbb{F}_1$ by, e.g., deploying PSV. However, minimizing \ac{MDT} or \ac{NDoT} is a much harder problem, let alone a simultaneous optimization of all three metrics.

Supporting a broad range of \ac{BP} sets has the benefit of high flexibility w.r.t.\@ the deployed data link technologies such as, e.g., \ac{MAC} protocols, sleeping patterns, etc. It is then possible to select the suitable set of \acp{BP} for each device, application, and deployment scenario. On the other hand, optimal or close-to-optimal discovery strategies allow for a more efficient resource allocation, smaller communication latency, and reduced energy consumption.

Based on our study of discovery strategies and the dependency of their performance on the structure of \ac{BP} sets, we would like to provide the following recommendations that may be useful for the development of new technologies and communication protocols for wireless communication that use periodic beacon messages for management or synchronization purposes or in case of deploying devices using existing technologies supporting a wide range of \acp{BP}, e.g. IEEE~802.11.

We argue that the best balance of flexibility and efficiency is provided by the family of \ac{BP} sets $\mathbb{F}_3$. It completely embraces and significantly extends the family of \ac{BP} sets supported by the IEEE~802.15.4 standard. For $\mathbb{F}_3$ there exist discovery strategies that are complete, and simultaneously optimize the \ac{WDT}, the \ac{MDT}, and the \ac{NDoT}. Moreover, these algorithms have a low complexity, which allows them to be executed on devices with constrained resources. 
An even much broader range of \ac{BP} sets is contained in the family of \ac{BP} sets $\mathbb{F}_2$. The higher flexibility comes at the cost of no longer being able to achieve optimality w.r.t.\@ the \ac{MDT} and the \ac{NDoT}. Nevertheless, for \ac{BP} sets from $\mathbb{F}_2$, it is still possible to achieve completeness and \ac{WDT}-optimality, as well as a close-to-optimal performance w.r.t.\@ \ac{MDT}, still with a low computational complexity.

We do not recommend to use the most general family of \ac{BP} sets $\mathbb{F}_1$. Even though our results show that the proposed algorithms still achieve remarkable performance over $\mathbb{F}_1$, we were not able to perform the evaluation with large settings, that is, with many channels and large \acp{BP} due to the high complexity of computing optimal \acp{MDT} that we are using as a performance benchmark. With large network environments, there will be specific cases that can result in considerably low performance of the discovery process independent of the deployed discovery approach.

\section{Conclusion}
\label{sec:conclusion}

In the present work we address the problem of asynchronous passive multi-channel discovery of neighbors periodically transmitting beacon messages. Our goal has been to develop approaches that guarantee a complete discovery, minimize \ac{WDT}, minimize  \ac{MDT}, and maximize  \ac{NDoT}. We aimed at designing solutions that give the device maximum flexibility in selecting its beaconing period, in order to optimally support its state, operational goals, and data link protocols.

We have completely characterized the class of schedules, we call them recursive, that are complete and optimized w.r.t.\@ the three targeted objectives. We have developed algorithms that, under certain assumptions, generate recursive schedules, while they are still applicable for the most general cases, where they exhibit optimal or close-to-optimal performance. Moreover, they significantly outperform the Passive Scan defined by the IEEE~802.15.4 standard. The proposed approaches can be used both for offline and online computations.

In addition, we have developed an \ac{ILP}-based approach minimizing the \ac{MDT} for the most general case, that, however, exhibits a high computational complexity and memory consumption and is, therefore, only applicable for offline computation and for network environments of moderate size. 

Based on the gained insights we provide recommendations on the selection of \ac{BP} sets that are as non-restrictive as possible, but still allow for an efficient discovery process.

Our future work will focus on sharing gossip information about discovered neighbors among devices in their beacon messages. By incorporating this information into the computation of listening schedules, devices will be able to further speed up the discovery.

\acrodef{AP}{Access Point} \acrodefplural{AP}[AP's]{Access Points}
\acrodef{BP}{Beacon Period} \acrodefplural{BP}[BP's]{Beacon Periods}
\acrodef{BO}{Beacon Order} \acrodefplural{BO}[BO's]{Beacon Orders}
\acrodef{WDT}{Worst-Case Discovery Time} \acrodefplural{WDT}[WDT's]{Worst-Case Discovery Times}
\acrodef{SWDT}{Sample Worst-Case Discovery Time} \acrodefplural{SCDT}[SWDT's]{Sample Worst-Case Discovery Times}
\acrodef{DTN}{Delay Tolerant Network} \acrodefplural{DTN}[DTN's]{Delay Tolerant Networks}
\acrodef{GCD}{Greatest Common Divisor} 
\acrodef{IoT}{Internet of Things}
\acrodef{ILP}{Integer Linear Program} \acrodefplural{ILP}[ILP's]{Integer Linear Programs}
\acrodef{LCM}{Least Common Multiple} \acrodefplural{LCM}[LCM's]{Least Common Multiples}
\acrodef{LP}{Linear Program} \acrodefplural{LP}[LP's]{Linear Programs}
\acrodef{MAC}{Media Access Control}
\acrodef{CR}{Cognitive Radio}
\acrodef{PU}{Primary User} \acrodefplural{PU}[PU's]{Primary Users}
\acrodef{SU}{Secondary User}\acrodefplural{SU}[SU's]{Secondary Users}
\acrodef{CH}{Channel Hopping}
\acrodef{ISM}{Industrial, Scientific and Medical}
\acrodef{MDT}{Mean Discovery Time} \acrodefplural{MDT}[MDT's]{Mean Discovery Times}
\acrodef{NDoT}{Number of Discoveries over Time}
\acrodef{OFDM}{Orthogonal Frequency-Division Multiplexing}
\acrodef{PSV}{Passive Scan}
\acrodef{SMDT}{Sample Mean Discovery Time} \acrodefplural{SMDT}[SMDT's]{Sample Mean Discovery Times}
\acrodef{SNDoT}{Sample Number of Discoveries over Time}
\acrodef{TU}{Time Unit} \acrodefplural{TU}[TU's]{Time Units}
\acrodef{CDF}{Cumulative Distribution Function} \acrodefplural{CDF}[CDF's]{Cumulative Distribution Functions}
\acrodef{PDF}{Probability Distribution Function} \acrodefplural{PDF}[PDF's]{Probability Distribution Functions}
\acrodef{WLAN}{Wireless Local Area Network} \acrodefplural{WLAN}[WLAN's]{Wireless Local Area Networks} 
\bibliographystyle{IEEEtran}
\bibliography{references}

\vspace{-10mm}

\begin{IEEEbiography}[{\includegraphics[width=1.0in,height=1.25in,clip,keepaspectratio]{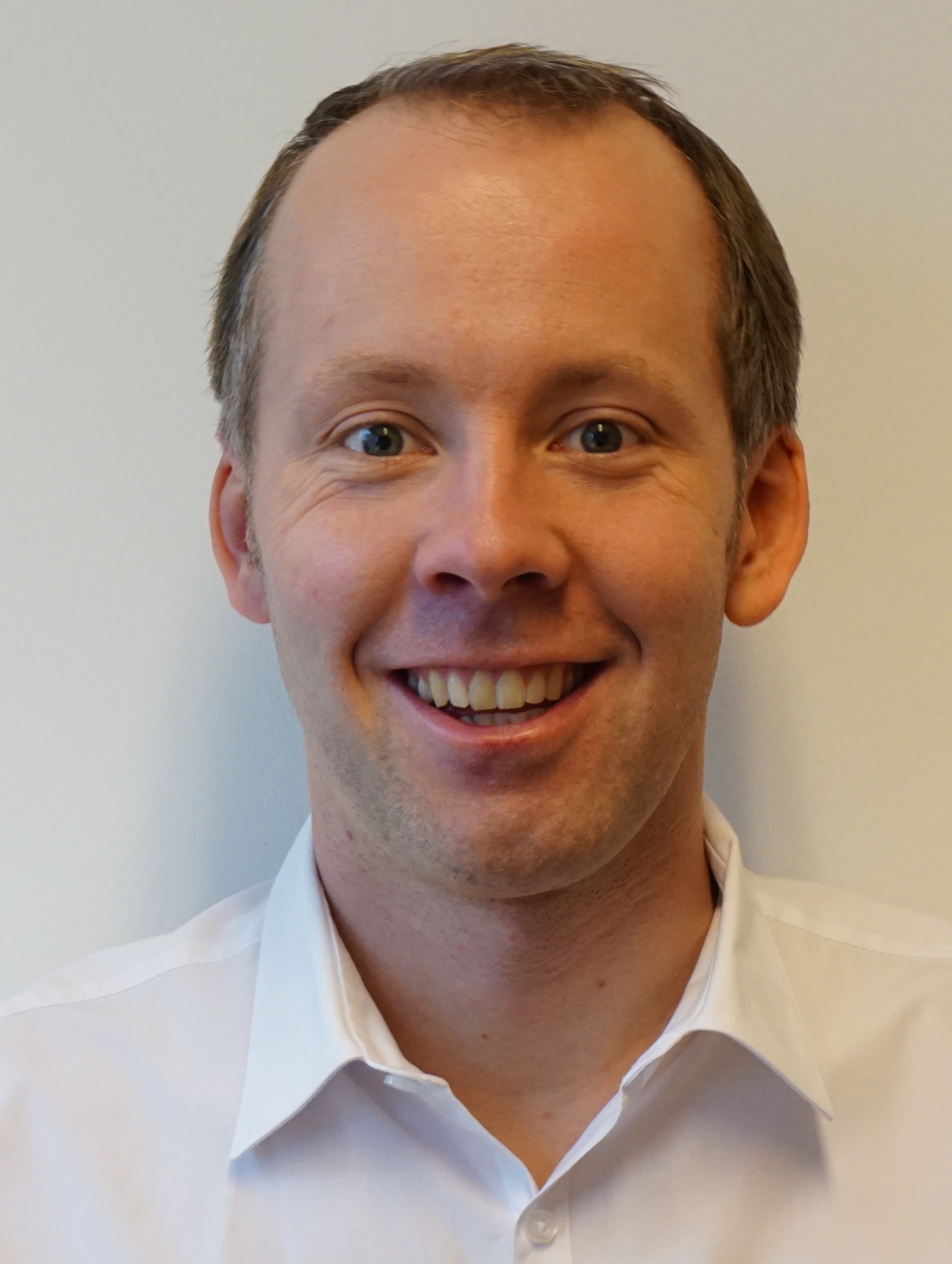}}]{Niels Karowski}
received a diploma degree in computer science in 2007 from the Technical University Berlin, Germany.
He is a PhD candidate at the Telecommunication Networks Group at Technical University of Berlin.
His research interests include wireless sensor networks, neighbor discovery and delay tolerant networks.
\end{IEEEbiography}

\vspace{-10mm}

\begin{IEEEbiography}[{\includegraphics[width=1.0in,height=1.25in,clip,keepaspectratio]{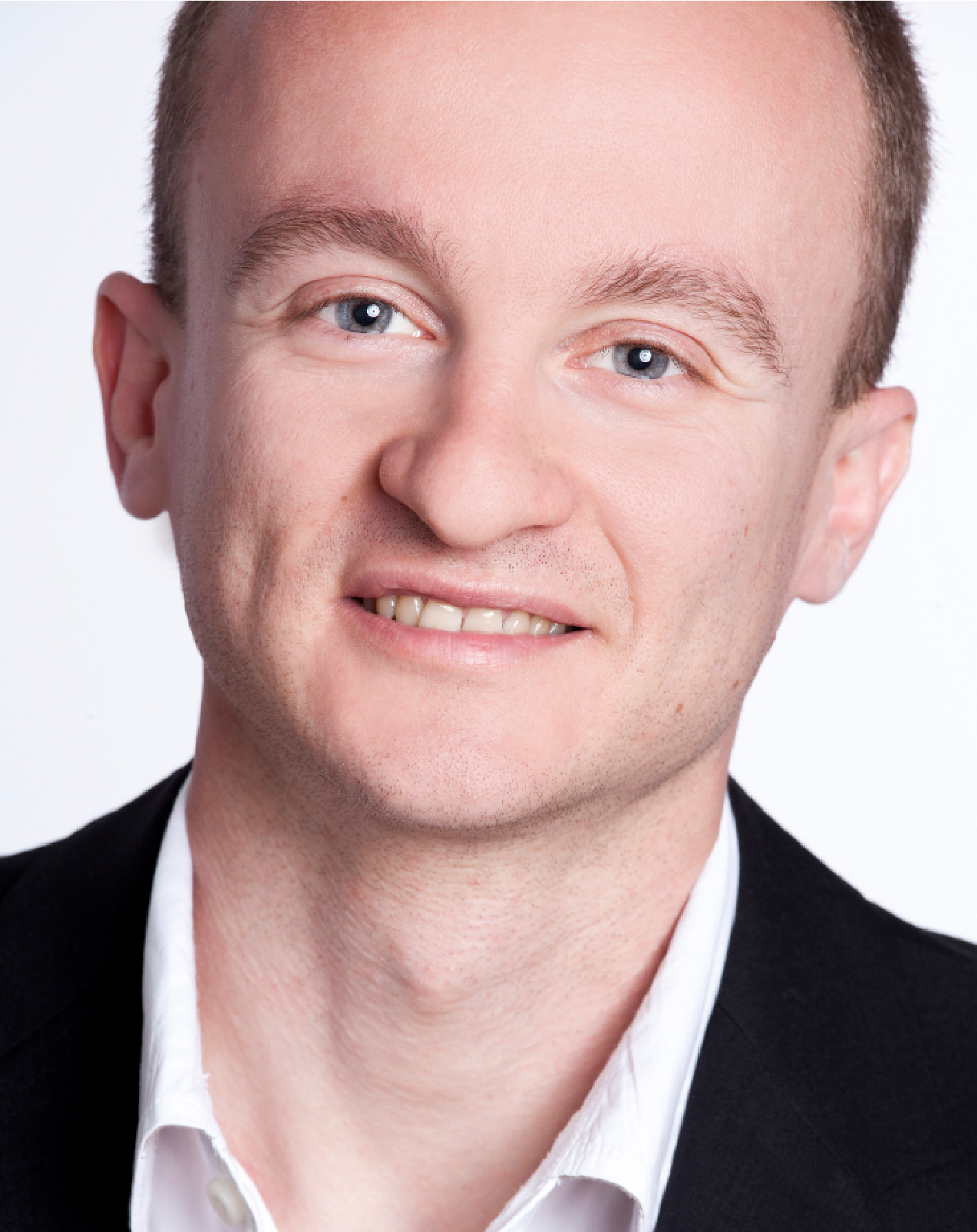}}]{Konstantin Miller}
Konstantin Miller received his diploma and Ph.D. degrees in Computer Engineering from the Technische Universit\"at Berlin, Germany, in 2007 and 2016. His diploma thesis received an award from the German Computer Science Society. He was awarded a Ph.D scholarship from the Innovation Center Human-Centric Communication at TUB, and has held research internships at STMicroelectronics in Milano, Italy, and at the University of Southern California in Los Angeles, CA, USA. His current research interests include multimedia streaming, wireless networks, and peer-to-peer networks, with a special focus on mathematical modeling.
\end{IEEEbiography}

\vspace{-10mm}

\begin{IEEEbiography}[{\includegraphics[width=1.0in,height=1.25in,clip,keepaspectratio]{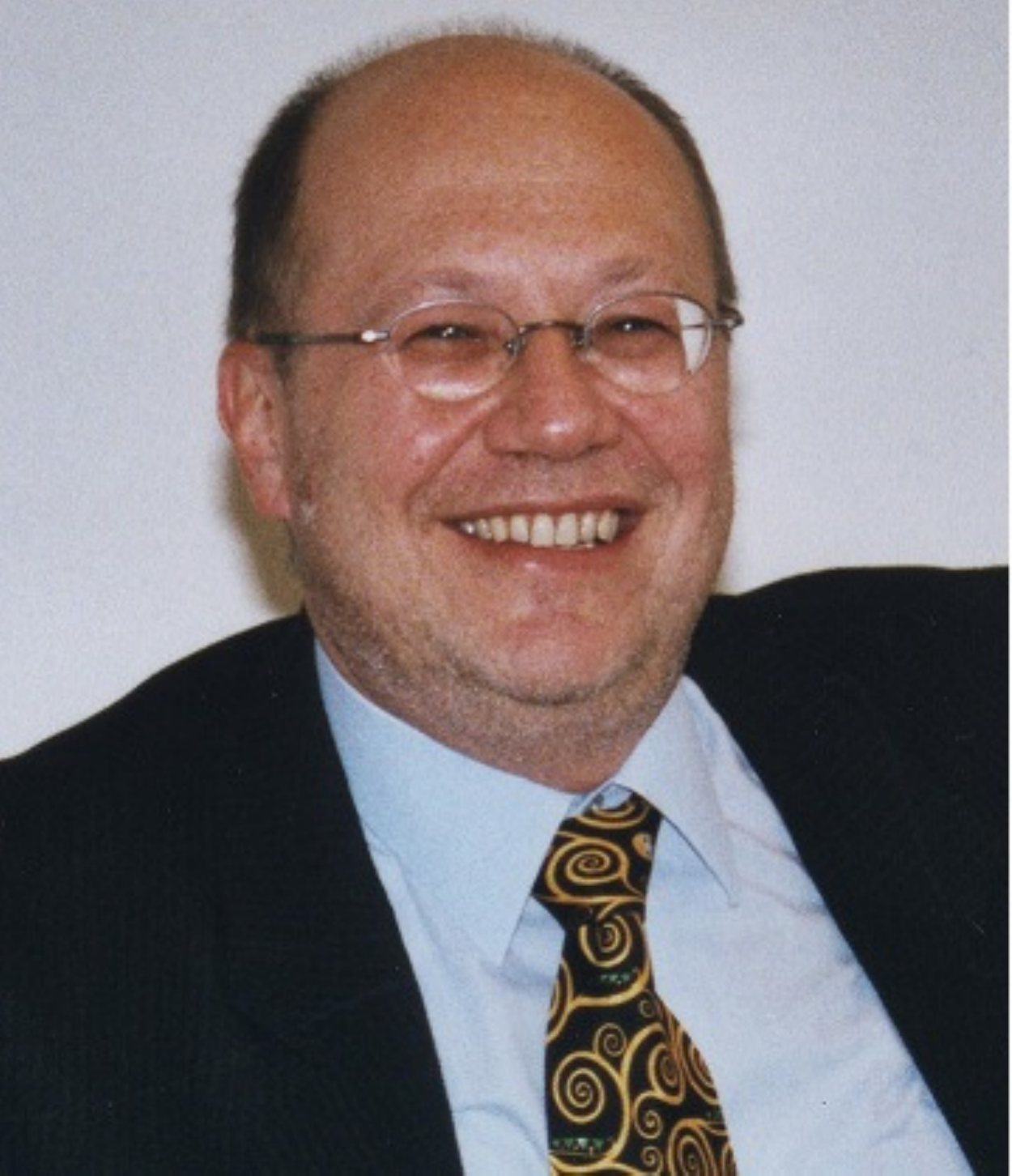}}]{Adam Wolisz}
received his degrees (Diploma 1972, Ph.D. 1976, Habil. 1983) from Silesian Unversity of Technology, Gliwice, Poland. He joined TU-Berlin in 1993, where he is a chaired professor in telecommunication networks and executive director of the Institute for Telecommunication Systems. He is also an adjunct professor at the Department of Electrical Engineering and Computer Science, University of California, Berkeley. His research interests are in architectures and protocols of communication networks. Recently he has been focusing mainly on wireless/mobile networking and sensor networks.
\end{IEEEbiography}

\cleardoublepage

\end{document}